\documentclass[journal, romanappendices]{IEEEtran}

\usepackage{graphicx}
\usepackage{amsmath}
\usepackage{balance}
\usepackage{cite}
\usepackage{url}

\usepackage{vmr-symbols-vecbold}
\usepackage{standard-macros}
\PassOptionsToPackage{hyphens}{url}
\usepackage{hyperref}

\DeclareSymbolFontAlphabet{\amsmathbb}{AMSb}%

\newcommand{\lro}[1]{\lefto({#1}\right)}																
\newcommand{\lrbo}[1]{\lefto \lbrace {#1} \right \rbrace}															
\newcommand{\lrho}[1]{\lefto [ {#1} \right ]}																				

\newcommand{\lr}[1]{\left({#1}\right)}																
\newcommand{\lrb}[1]{\left \lbrace {#1} \right \rbrace}															

\safemath{\dopplerspread}{B_D}																								
\safemath{\delayspread}{T_D}																									
\safemath{\nc}{n\sub{c}}																										
\safemath{\nf}{n\sub{f}}																										
\safemath{\efa}{p\sub{sc}}
\safemath{\efb}{p\sub{cs}}
\safemath{\ef}{\epsilon\sub{f}	}
\safemath{\nd}{n\sub{d}}																										
\safemath{\ntx}{n\sub{t}} 																											
\safemath{\nrx}{n\sub{r}}																											
\safemath{\ntxt}{\tilde{n\sub{t}}}																											
\safemath{\cb}{\ensuremath{L}} 																								
\safemath{\cl}{\ensuremath{n}} 																								
\safemath{\txanto}{{\ensuremath{\tilde{m}_t}}} 																		
\safemath{\cs}{M} 																														
\safemath{\idPustm}{\ensuremath{S_{k}}}
\safemath{\error}{\ensuremath{\epsilon}} 																				
\safemath{\eexp}{\ensuremath{\mathcal{E}}} 																			
\safemath{\nsubc}{n\sub{s}}			 																						
\safemath{\nofdm}{n\sub{o}} 																									
\safemath{\bc}{\ensuremath{B_c}} 																							
\safemath{\ts}{\ensuremath{T_s}} 																							
\safemath{\nrb}{\ensuremath{n_{rb}}} 																						
\safemath{\rul}{\ensuremath{\rho\sub{ul}}}
\safemath{\rdl}{\ensuremath{\rho\sub{dl}}}

\safemath{\nres}{\ell}
\safemath{\nr}{n\sub{r}}
\safemath{\maxk}{M^*\lr{\nres, \nsubc, \nofdm, \epsilon, \rho}}
\safemath{\Rmax}{R^*}
\safemath{\Emin}{E\sub{b}^*/N_0}
\safemath{\Eminf}{\frac{E\sub{b}^*}{N_0}}
\safemath{\np}{\ensuremath{n\sub{p}}}
\safemath{\ndf}{\ensuremath{\bar{n}\sub{d}}}
\safemath{\npf}{\ensuremath{\bar{n}\sub{p}}}
\safemath{\code}{\ensuremath{\mathcal{C}}}
\safemath{\err}{\ensuremath{\epsilon}}
\safemath{\rp}{\ensuremath{\rho\sub{p}}}
\safemath{\rd}{\ensuremath{\rho\sub{d}}}
\safemath{\cohtime}{\ensuremath{T\sub{c}}}
\safemath{\cohbw}{\ensuremath{B\sub{c}}}
\safemath{\nmax}{\ensuremath{\ell\sub{m}}}
\safemath{\ntot}{\ensuremath{n\sub{tot}}}
\safemath{\nul}{\ensuremath{n\sub{ul}}}
\safemath{\ndl}{\ensuremath{n\sub{dl}}}

\safemath{\yp}{\ensuremath{\randvecy_{\nu}^{(\text{p})}}}
\safemath{\yd}{\ensuremath{\randvecy_{\nu}^{(\text{d})}}}
\safemath{\ypd}{\ensuremath{\vecy_{\nu}^{(\text{p})}}}
\safemath{\ydd}{\ensuremath{\vecy_{\nu}^{(\text{d})}}}

\safemath{\ypf}{\ensuremath{\bar{\randvecy}_{\nu}^{(\text{p})}}}
\safemath{\ydf}{\ensuremath{\bar{\randvecy}_{\nu}^{(\text{d})}}}
\safemath{\ypdf}{\ensuremath{\bar{\vecy}_{\nu}^{(\text{p})}}}
\safemath{\yddf}{\ensuremath{\bar{\vecy}_{\nu}^{(\text{d})}}}

\safemath{\xp}{\ensuremath{\vecx^{(\text{p})}}}
\safemath{\xd}{\ensuremath{\randvecx_{\nu}^{(\text{d})}}}
\safemath{\xdd}{\ensuremath{\vecx_{\nu}^{(\text{d})}}}

\safemath{\xpf}{\ensuremath{\bar{\vecx}^{(\text{p})}}}
\safemath{\xdf}{\ensuremath{\bar{\randvecx}_{\nu}^{(\text{d})}}}
\safemath{\xddf}{\ensuremath{\bar{\vecx}_{\nu}^{(\text{d})}}}

\safemath{\xdb}{\ensuremath{\overline{\randvecx}^{(\text{d})}}}
\safemath{\Pxd}{\ensuremath{P_{\randvecx^{(\text{d})}}}}

\safemath{\xpbar}{\ensuremath{\overline{\matX}^{(\text{p})}}}
\safemath{\xdbar}{\ensuremath{\overline{\randmatX}^{(\text{d})}}}

\safemath{\xdv}{\ensuremath{\randvecx^{(\text{d})}}}
\safemath{\xdbarv}{\ensuremath{\overline{\randvecx}^{(\text{d})}}}
\safemath{\ydv}{\ensuremath{\randvecy^{(\text{d})}}}

\safemath{\xdr}{\ensuremath{\matX^{(\text{d})}}}

\safemath{\ttx}{\ensuremath{\tau\sub{tx}}}
\safemath{\trx}{\ensuremath{\tau\sub{rx}}}
\safemath{\ack}{\ensuremath{\mathrm{s}}}
\safemath{\nack}{\ensuremath{\mathrm{c}}}

\newcommand{\prob}[1]{\ensuremath{\mathbb{P}\lrho{#1}}}

\safemath{\mI}{\ensuremath{i\lro{\randvecy ; \randvecx}}} 				

\safemath{\randveca}{\bm{A}}
\safemath{\randvecb}{\bm{B}}
\safemath{\randvecc}{\bm{C}}
\safemath{\randvecd}{\bm{D}}
\safemath{\randvece}{\bm{E}}
\safemath{\randvecf}{\bm{F}}
\safemath{\randvecg}{\bm{G}}
\safemath{\randvech}{\bm{H}}
\safemath{\randveci}{\bm{I}}
\safemath{\randvecj}{\bm{J}}
\safemath{\randveck}{\bm{K}}
\safemath{\randvecl}{\bm{L}}
\safemath{\randvecm}{\bm{M}}
\safemath{\randvecn}{\bm{N}}
\safemath{\randveco}{\bm{O}}
\safemath{\randvecp}{\bm{P}}
\safemath{\randvecq}{\bm{Q}}
\safemath{\randvecr}{\bm{R}}
\safemath{\randvecs}{\bm{S}}
\safemath{\randvect}{\bm{T}}
\safemath{\randvecu}{\bm{U}}
\safemath{\randvecv}{\bm{V}}
\safemath{\randvecw}{\bm{W}}
\safemath{\randvecx}{\bm{X}}
\safemath{\randvecy}{\bm{Y}}
\safemath{\randvecz}{\bm{Z}}
\safemath{\randvecphi}{\bm{\Phi}}

\safemath{\randmatA}{\amsmathbb{A}}
\safemath{\randmatB}{\amsmathbb{B}}
\safemath{\randmatC}{\amsmathbb{C}}
\safemath{\randmatD}{\amsmathbb{D}}
\safemath{\randmatE}{\amsmathbb{E}}
\safemath{\randmatF}{\amsmathbb{F}}
\safemath{\randmatG}{\amsmathbb{G}}
\safemath{\randmatH}{\amsmathbb{H}}
\safemath{\randmatI}{\amsmathbb{I}}
\safemath{\randmatJ}{\amsmathbb{J}}
\safemath{\randmatK}{\amsmathbb{K}}
\safemath{\randmatL}{\amsmathbb{L}}
\safemath{\randmatM}{\amsmathbb{M}}
\safemath{\randmatN}{\amsmathbb{N}}
\safemath{\randmatO}{\amsmathbb{O}}
\safemath{\randmatP}{\amsmathbb{P}}
\safemath{\randmatQ}{\amsmathbb{Q}}
\safemath{\randmatR}{\amsmathbb{R}}
\safemath{\randmatS}{\amsmathbb{S}}
\safemath{\randmatT}{\amsmathbb{T}}
\safemath{\randmatU}{\amsmathbb{U}}
\safemath{\randmatV}{\amsmathbb{V}}
\safemath{\randmatW}{\amsmathbb{W}}
\safemath{\randmatX}{\amsmathbb{X}}
\safemath{\randmatY}{\amsmathbb{Y}}
\safemath{\randmatZ}{\amsmathbb{Z}}
\safemath{\randmatSigma}{\mathbb{\Sigma}}
\safemath{\randmatPhi}{\mathbb{\Phi}}
\safemath{\randmatLambda}{\mathbb{\Lambda}}

\safemath{\matSigma}{\bm{\Sigma}}
\safemath{\matPhi}{\bm{\Phi}}
\safemath{\matLambda}{\bm{\Lambda}}

\usepackage{glossaries}
\glsdisablehyper
\newglossaryentry{simo}
{
  name={SIMO},
  description={single-input multiple-output},
  first={\glsentrydesc{simo} (\glsentrytext{simo})}
}

\newglossaryentry{mmse}
{
  name={MMSE},
  description={minimum mean-square error},
  first={\glsentrydesc{mmse} (\glsentrytext{mmse})}
}

\newglossaryentry{na}
{
  name={NA},
  description={normal approximation},
  first={\glsentrydesc{na} (\glsentrytext{na})},
  plural={NAs},
  descriptionplural={normal approximations},
  firstplural={\glsentrydescplural{na} (\glsentryplural{na})}
}

\newglossaryentry{zf}
{
  name={ZF},
  description={zero-forcing},
  first={\glsentrydesc{zf} (\glsentrytext{zf})}
}

\newglossaryentry{sir}
{
  name={SIR},
  description={signal-to-interference ratio},
  first={\glsentrydesc{sir} (\glsentrytext{sir})}
}

\newglossaryentry{mse}
{
  name={MSE},
  description={mean-square error},
  first={\glsentrydesc{mse} (\glsentrytext{mse})}
}
\newglossaryentry{mmmse}
{
  name={M-MMSE},
  description={multicell MMSE},
  first={\glsentrydesc{mmmse} (\glsentrytext{mmmse})}
}

\newglossaryentry{ls}
{
  name={LS},
  description={least-squares},
  first={\glsentrydesc{ls} (\glsentrytext{ls})}
}

\newglossaryentry{mr}
{
  name={MR},
  description={maximum ratio},
  first={\glsentrydesc{mr} (\glsentrytext{mr})}
}

\newglossaryentry{ue}
{
  name={UE},
  description={user equipment},
  first={\glsentrydesc{ue} (\glsentrytext{ue})},
  plural={UEs},
  descriptionplural={user equipments},
  firstplural={\glsentrydescplural{ue} (\glsentryplural{ue})}
}

\newglossaryentry{bs}
{
  name={BS},
  description={base station},
  first={\glsentrydesc{bs} (\glsentrytext{bs})},
  plural={BS},
  descriptionplural={base stations},
  firstplural={\glsentrydescplural{bs} (\glsentryplural{bs})}
}

\newglossaryentry{ostbc}
{
  name={OSTBC},
  description={orthogonal space-time block code},
  first={\glsentrydesc{ostbc} (\glsentrytext{ostbc})},
  plural={OSTBCs},
  descriptionplural={orthogonal space-time block codes},
  firstplural={\glsentrydescplural{ostbc} (\glsentryplural{ostbc})}
}

\newglossaryentry{biawgn}
{
  name={bi-AWGN},
  description={binary-input additive white Gaussian noise},
  first={\glsentrydesc{biawgn} (\glsentrytext{biawgn})}
}

\newglossaryentry{flnf}
{
  name={FLNF},
  description={fixed-length no-feedback},
  first={\glsentrydesc{flnf} (\glsentrytext{flnf})}
}

\newglossaryentry{crc}
{
  name={CRC},
  description={cyclic redundancy check},
  first={\glsentrydesc{crc} (\glsentrytext{crc})}
}

\newglossaryentry{bsc}
{
  name={BSC},
  description={binary symmetric channel},
  first={\glsentrydesc{bsc} (\glsentrytext{bsc})}
  plural={BSCs},
  descriptionplural={binary symmetric channels},
  firstplural={\glsentrydescplural{bsc} (\glsentryplural{bsc})}
}

\newglossaryentry{bec}
{
  name={BEC},
  description={binary-erasure channel},
  first={\glsentrydesc{bec} (\glsentrytext{bec})}
}
\newglossaryentry{awgn}
{
  name={AWGN},
  description={additive white Gaussian noise},
  first={\glsentrydesc{awgn} (\glsentrytext{awgn})}
}
\newglossaryentry{3gpp}
{
  name={3GPP},
  description={3rd generation partnership project},
  first={\glsentrydesc{3gpp} (\glsentrytext{3gpp})}
}

\newglossaryentry{dl}
{
  name={DL},
  description={downlink},
  first={\glsentrydesc{dl} (\glsentrytext{dl})}
}

\newglossaryentry{LED}
{
  name={LED},
  description={light emitting diode},
  first={\glsentrydesc{LED} (\glsentrytext{LED})},
  plural={LEDs},
  descriptionplural={light emitting diodes},
  firstplural={\glsentrydescplural{LED} (\glsentryplural{LED})}
}

\newglossaryentry{lte}
{
  name={LTE},
  description={long term evolution},
  first={\glsentrydesc{lte} (\glsentrytext{lte})}
}
\newglossaryentry{ofdm}
{
  name={OFDM},
  description={orthogonal frequency-division multiplexing},
  first={\glsentrydesc{ofdm} (\glsentrytext{ofdm})}
}

\newglossaryentry{ul}
{
  name={UL},
  description={uplink},
  first={\glsentrydesc{ul} (\glsentrytext{ul})}
}
\newglossaryentry{rb}
{
  name={RB},
  description={resource block},
  first={\glsentrydesc{rb} (\glsentrytext{rb})},
  plural={RBs},
  descriptionplural={resource blocks},
  firstplural={\glsentrydescplural{rb} (\glsentryplural{rb})}
}
\newglossaryentry{cu}
{
  name={CU},
  description={channel use},
  first={\glsentrydesc{cu} (\glsentrytext{cu})},
  plural={CUs},
  descriptionplural={channel uses},
  firstplural={\glsentrydescplural{cu} (\glsentryplural{cu})}
}
\newglossaryentry{tti}
{
  name={TTI},
  description={transmission time interval},
  first={\glsentrydesc{tti} (\glsentrytext{tti})},
  plural={TTI's},
  descriptionplural={transmission time intervals},
  firstplural={\glsentrydescplural{tti} (\glsentryplural{tti})}
}
\newglossaryentry{iid}
{
  name={i.i.d.},
  description={independent and identically distributed},
  first={\glsentrydesc{iid} (\glsentrytext{iid})}
}
\newglossaryentry{psd}
{
  name={PSD},
  description={power spectral density},
  first={\glsentrydesc{psd} (\glsentrytext{psd})}
}

\newglossaryentry{mimo}
{
  name={MIMO},
  description={multiple-input multiple-output},
  first={\glsentrydesc{mimo} (\glsentrytext{mimo})}
}

\newglossaryentry{bler}
{
  name={BLER},
  description={block error probability},
  first={\glsentrydesc{bler} (\glsentrytext{bler})}
}
\newglossaryentry{mtc}
{
  name={MTC},
  description={machine-type communication},
  first={\glsentrydesc{mtc} (\glsentrytext{mtc})}
}
\newacronym{csi}{CSI}{channel state information}
\newglossaryentry{dvbt}
{
  name={DVB-T},
  description={terrestrial digital video broadcasting},
  first={\glsentrydesc{dvbt} (\glsentrytext{dvbt})}
}
\newglossaryentry{pat}
{
  name={PAT},
  description={pilot-assisted transmission},
  first={\glsentrydesc{pat} (\glsentrytext{pat})}
}
\newglossaryentry{ml}
{
  name={ML},
  description={maximum likelihood},
  first={\glsentrydesc{ml} (\glsentrytext{ml})}
}
\newglossaryentry{dt}
{
  name={DT},
  description={dependence-testing},
  first={\glsentrydesc{dt} (\glsentrytext{dt})}
}
\newglossaryentry{ustm}
{
  name={USTM},
  description={unitary space-time modulation},
  first={\glsentrydesc{ustm} (\glsentrytext{ustm})}
}
\newglossaryentry{siso}
{
  name={SISO},
  description={single-input single-output},
  first={\glsentrydesc{siso} (\glsentrytext{siso})}
}
\newglossaryentry{snn}
{
  name={SNN},
  description={scaled nearest-neighbor},
  first={\glsentrydesc{snn} (\glsentrytext{snn})}
}

\newglossaryentry{snnd}
{
  name={SNND},
  description={scaled nearest-neighbor decoding},
  first={\glsentrydesc{snnd} (\glsentrytext{snnd})}
}
\newglossaryentry{rcus}
{
  name={RCUs},
  description={random-coding union bound with parameter $s$},
  first={\glsentrydesc{rcus} (\glsentrytext{rcus})}
}
\newglossaryentry{osd}
{
  name={OSD},
  description={ordered statistics decoding},
  first={\glsentrydesc{osd} (\glsentrytext{osd})}
}
\newglossaryentry{llr}
{
  name={LLR},
  description={log-likelihood ratio},
  first={\glsentrydesc{llr} (\glsentrytext{llr})}
   plural={LLR's},
  descriptionplural={log-likelihood ratios},
  firstplural={\glsentrydescplural{llr} (\glsentryplural{llr})}
}
\newglossaryentry{qpsk}
{
  name={QPSK},
  description={quaternary phase shift keying},
  first={\glsentrydesc{qpsk} (\glsentrytext{qpsk})}
}
\newglossaryentry{nlos}
{
  name={NLOS},
  description={non-line-of-sight},
  first={\glsentrydesc{nlos} (\glsentrytext{nlos})}
}
\newglossaryentry{los}
{
  name={LOS},
  description={line-of-sight},
  first={\glsentrydesc{los} (\glsentrytext{los})}
}
\newglossaryentry{mgf}
{
  name={MGF},
  description={moment-generating function},
  first={\glsentrydesc{mgf} (\glsentrytext{mgf})}
}
\newglossaryentry{cgf}
{
  name={CGF},
  description={cumulant-generating function},
  first={\glsentrydesc{cgf} (\glsentrytext{cgf})},
  plural={CGFs},
  descriptionplural={cumulant-generating functions},
  firstplural={\glsentrydescplural{cgf} (\glsentryplural{cgf})}
}
\newglossaryentry{pdf}
{
  name={PDF},
  description={probability density function},
  first={\glsentrydesc{pdf} (\glsentrytext{pdf})},
   plural={PDF's},
  descriptionplural={probability density functions},
  firstplural={\glsentrydescplural{pdf} (\glsentryplural{pdf})}
}

\newglossaryentry{ccdf}
{
  name={CCDF},
  description={complementary cumulative distribution function},
  first={\glsentrydesc{ccdf} (\glsentrytext{ccdf})}
}

\newglossaryentry{cdf}
{
  name={CDF},
  description={cumulative distribution function},
  first={\glsentrydesc{cdf} (\glsentrytext{cdf})}
}

\newglossaryentry{gmi}
{
  name={GMI},
  description={generalized mutual information},
  first={\glsentrydesc{gmi} (\glsentrytext{gmi})}
}

\newglossaryentry{arq}
{
  name={ARQ},
  description={automatic repeat request},
  first={\glsentrydesc{arq} (\glsentrytext{arq})}
}

\newglossaryentry{harq}
{
  name={HARQ},
  description={hybrid automatic repeat request},
  first={\glsentrydesc{harq} (\glsentrytext{harq})}
}

\newglossaryentry{fbl}
{
  name={FBL-NF},
  description={fixed blocklength with no feedback},
  first={\glsentrydesc{fbl} (\glsentrytext{fbl})}
}
\newglossaryentry{ssnf}
{
  name={SS-NF},
  description={single-shot no-feedback},
  first={\glsentrydesc{ssnf} (\glsentrytext{ssnf})}
}

\newglossaryentry{urllc}
{
  name={URLLC},
  description={ultra-reliable low-latency communications},
  first={\glsentrydesc{urllc} (\glsentrytext{urllc})}
}

\newglossaryentry{vlsf}
{
  name={VLSF},
  description={Variable-length stop-feedback},
  first={\glsentrydesc{vlsf} (\glsentrytext{vlsf})}
}

\newglossaryentry{vlnsf}
{
  name={VLNSF},
  description={variable-length noisy stop feedback},
  first={\glsentrydesc{vlnsf} (\glsentrytext{vlnsf})}
}

\newglossaryentry{dmt}
{
  name={DMT},
  description={diversity-multiplexing tradeoff},
  first={\glsentrydesc{dmt} (\glsentrytext{dmt})}
}

\newglossaryentry{dmdt}
{
  name={DMDT},
  description={diversity-multiplexing-delay tradeoff},
  first={\glsentrydesc{dmdt} (\glsentrytext{dmdt})}
}

\newglossaryentry{tddt}
{
  name={TDDT},
  description={throughput-diversity-delay tradeoff},
  first={\glsentrydesc{tddt} (\glsentrytext{tddt})}
}

\newglossaryentry{tdd}
{
  name={TDD},
  description={time-division duplex},
  first={\glsentrydesc{tdd} (\glsentrytext{tdd})}
}

\newglossaryentry{stbc}
{
  name={STBC},
  description={space-time block-codes},
  first={\glsentrydesc{stbc} (\glsentrytext{stbc})},
  plural={STBC's},
  descriptionplural={space-time block-codes},
  firstplural={\glsentrydescplural{stbc} (\glsentryplural{stbc})}
}

\newglossaryentry{harqir}
{
  name={HARQ-IR},
  description={hybrid automatic repeat request with incremental redundancy},
  first={\glsentrydesc{harqir} (\glsentrytext{harqir})}
}

\newglossaryentry{harqcc}
{
  name={HARQ-CC},
  description={hybrid automatic repeat request with chase combining},
  first={\glsentrydesc{harqcc} (\glsentrytext{harqcc})}
}

\newglossaryentry{wp1}
{
  name={w.p.1.},
  description={with probability one},
  first={\glsentrydesc{wp1} (\glsentrytext{wp1})}
}

\newglossaryentry{vlff}
{
  name={VLFF},
  description={variable-length full feedback},
  first={\glsentrydesc{vlff} (\glsentrytext{vlff})}
}

\newglossaryentry{dmc}
{
  name={DMC},
  description={discrete memoryless channel},
  first={\glsentrydesc{dmc} (\glsentrytext{dmc})}
  plural={DMCs},
  descriptionplural={discrete memoryless channels},
  firstplural={\glsentrydescplural{dmc} (\glsentryplural{dmc})}
}

\usepackage{pgfplots}
\pgfplotsset{compat=1.17}
\usepgfplotslibrary{groupplots}
\usetikzlibrary{pgfplots.groupplots} 
\usepackage{subcaption}
\let\abs\undefined
\newcommand{\abs}[1]{\lvert#1\rvert}		

\def\tr{\mathrm{tr}}
\def\trace{\mathrm{tr}}

\newtheorem{theorem}{Theorem}
\newtheorem{lemma}{Lemma}

\newtheorem{assumption}{Assumption}

\def\tr{\mathrm{tr}}

\def\tr{\mathrm{tr}}

\def\Htran{\mbox{\tiny $\mathrm{H}$}}
\def\Ttran{\mbox{\tiny $\mathrm{T}$}}
\def\bphiu{\boldsymbol{\phi}} 

\def\Qexp{\Psi_{n,\zeta}}

\begin{document}

\title{URLLC with Massive MIMO:\\ Analysis and Design at Finite Blocklength \vspace{-0.1cm}}

\author{Johan \"Ostman,~\IEEEmembership{Student Member,~IEEE}, Alejandro Lancho,~\IEEEmembership{Member,~IEEE}, Giuseppe~Durisi,~\IEEEmembership{Senior Member,~IEEE}, and Luca Sanguinetti,~\IEEEmembership{Senior Member,~IEEE} \vspace{-0.6cm}

\thanks{Parts of this paper have been presented at the Asilomar Conf. Signals, Syst., Comput., Pacific Grove, CA, USA, Dec. 2019~\cite{ostman19-12}, and will be presented at the IEEE Int. Conf. Commun. (ICC), Montreal, Canada, Jun. 2021.}

\thanks{Johan \"Ostman, Alejandro Lancho, and Giuseppe Durisi are with the Department of Electrical Engineering, Chalmers University of Technology, Gothenburg 41296, Sweden (e-mail: \{johanos,lanchoa,durisi\}@chalmers.se). Luca Sanguinetti is with the Dipartimento di Ingegneria dell'Informazione, University of Pisa, 56122 Pisa, Italy (e-mail: luca.sanguinetti@unipi.it).
\newline\indent The work of Johan Östman, Alejandro Lancho and Giuseppe Durisi was partly supported by the Swedish Research Council under grant 2016-03293, and by the Wallenberg AI, Autonomous Systems, and Software Program. Luca Sanguinetti was in part supported by the Italian Ministry of Education and Research (MIUR) in the framework of the CrossLab project (Departments of Excellence).
}
}
 \maketitle
  \sloppy 
\begin{abstract}
The fast adoption of Massive MIMO for high-throughput communications was enabled by many research contributions mostly relying on infinite-blocklength information-theoretic bounds. 
This makes it hard to assess the suitability of Massive MIMO for ultra-reliable low-latency communications (URLLC) operating with short-blocklength codes. 
This paper provides a rigorous framework for the characterization and numerical  evaluation (using the saddlepoint approximation) of the error probability achievable in the uplink and downlink of Massive MIMO at finite blocklength.
The framework encompasses imperfect channel state information, pilot contamination, spatially correlated channels, and arbitrary linear spatial processing. 
In line with previous results based on infinite-blocklength bounds, we prove that, with minimum mean-square error (MMSE) processing and spatially correlated channels, the error probability at finite blocklength goes to zero as the number $M$ of antennas grows to infinity, even under pilot contamination. 
However, numerical results for a practical URLLC network setup involving a base station with  $M=100$ antennas, show that a target error probability of $10^{-5}$ can be achieved with MMSE processing, uniformly over each cell, only if orthogonal pilot sequences are assigned to all the users in the network. Maximum ratio processing does not suffice.
\end{abstract}


 
\section{Introduction}\label{sec:intro}
Among the new use cases that will be supported by next generation wireless systems~\cite{3GPPTS38.300},
some of the most challenging ones fall into the category of \gls{urllc}.  
For example, in \gls{urllc} for factory automation~\cite{3GPP22.104}, small payloads on the order of $100$ bits must be delivered within hundreds of microseconds and with a reliability no smaller than  $99.999\%$.
To achieve such a high reliability, it is crucial to exploit diversity. Unfortunately, the stringent latency requirements prevent the exploitation of diversity in time.
Furthermore, the use of frequency diversity is problematic, especially in the uplink where current standardization rules do not allow \glspl{ue} to spread a packet over independently fading frequency resources. 
Thus, the spatial diversity offered by multiple antennas becomes critical to achieve the desired reliability. The latest instantiation of multiple antenna technologies is the so-called Massive MIMO (multiple-input multiple-output), which refers to a wireless network where \glspl{bs}, equipped with a very large number $M$ of antennas, serve a multitude of \glspl{ue} via linear spatial signal processing~\cite{Marzetta10}. 
Thanks to the intense research performed since its inception in 2010, 
the advantages of Massive MIMO in terms of spectral efficiency~\cite{Ngo13,Sanguinetti18}, energy efficiency~\cite{bjornson15-a}, and power control~\cite{Cheng17} are well understood, and its key ingredients have made it into the 5G standard~\cite{bjornson19-b}. 
However, all these results have mainly been established in the \textit{ergodic regime}, where the propagation channel evolves according to a block-fading model, and each codeword spans an increasingly large number of independent fading realizations as the codeword length goes to infinity (infinite-blocklength regime).  
Since these assumptions are highly questionable in \gls{urllc} scenarios~\cite{durisi16-09a}, it remains unclear  whether the design guidelines that have been obtained so far for Massive MIMO (see~\cite{bjornson19,marzetta16-a} for a detailed review on the topic) apply to \gls{urllc} deployments.

\subsection{Prior Art}
Unlike the vast majority of literature on Massive MIMO, which focuses on the aforementioned ergodic regime, the authors in~\cite{Karlsson18,bana18-10a} assume that the fading channel stays constant during the transmission of a codeword (the so-called quasi-static fading scenario) and use outage capacity~\cite{ozarow94-05a} as asymptotic performance metric.
Although the quasi-static fading scenario is relevant for \gls{urllc}, the infinite blocklength assumption may yield incorrect estimates of the error probability.
The use of outage capacity in the context of \gls{urllc} is often justified by the results reported in~\cite{yang14-07c}, where it is proved that short channel codes operate close to the outage capacity for quasi-static fading channels.
More specifically, the authors of~\cite{yang14-07c} proved that the difference between the outage capacity and the maximum coding rate, achievable at finite blocklength over quasi-static fading channels, goes to zero much faster than the difference between the capacity and the maximum coding rate achievable over \gls{awgn} channels.
The intuition is that the dominant sources of errors in quasi-static fading channels are deep-fade events, which cannot be alleviated through the use of channel codes, since channel coding provides protection only against additive noise.

The application of this result to Massive MIMO is problematic since, as $M$ grows, we start observing channel hardening and the underlying effective channel (after precoding/combining) becomes more similar to an \gls{awgn} channel.
As a consequence, finite-blocklength effects become more pronounced, since additive noise turns into the dominating impairment.
Another unsatisfactory feature of the outage-capacity framework is its inability to account for the \gls{csi} acquisition overhead, caused by the transmission of pilot sequences.
Indeed, quasi-static fading channels can be learnt perfectly at the receiver in the asymptotic limit of large blocklength with no rate penalty: it is enough to let the number of pilot symbols grow sublinearly with the blocklength. 
The attempts made so far to include channel-estimation overhead in the outage setup~\cite{Karlsson18,bana18-10a} are not convincing from a theoretical perspective.
A theoretically satisfying framework must include the use of a mismatch receiver that treats the channel estimate, obtained using a fixed number of pilot symbols, as perfect.
One difficulty is that a fundamental result commonly used in the ergodic case to bound the mutual information, by treating the channel estimation error as noise (see, e.g., \cite[Lemma~B.0.1]{lapidoth02-05a}), does not apply to the outage case.
This is because, in the outage setup, the fading channel stays constant over the entire codeword, and one is interested in computing an outage event over fading realizations. This  means that both the channel and its estimate must be treated as deterministic quantities when computing bounds on the instantaneous spectral efficiency. 

The limitation of both ergodic and outage setups can be overcome by performing a nonasymptotic analysis of the error probability based on the finite-blocklength information-theoretic bounds introduced in~\cite{polyanskiy10-05a} and extended to fading channels in~\cite{yang14-07c,durisi16-02a,ostman19-02}.
This approach has been pursued recently in~\cite{Zeng2020, Ren2020}.
However, the analysis in these papers relies on the so called \textit{normal approximation}~\cite[Eq.~(291)]{polyanskiy10-05a}, whose tightness for the range of error probabilities of interest in \gls{urllc} is questionable.  
Also, the use of the normal approximation for the case of imperfect \gls{csi} in both~\cite{Zeng2020, Ren2020} is not convincing, since the approximation does not depend on the instantaneous channel estimation error, but only on its variance.
This is not compatible with a scenario in which the channel stays constant over the duration of each codeword.  
\subsection{Contributions}
To verify if the design guidelines developed for Massive MIMO in the context of non-delay limited, large-throughput, communication links apply also to the \gls{urllc} setup, we present a rigorous nonasymptotic characterization of the error probability achievable in Massive MIMO.
Specifically, we provide a firm upper bound on the error probability, which is obtained by adapting the \gls{rcus}  introduced in~\cite{martinez11-02a} to the case of Massive MIMO communications.
The resulting bound applies to Gaussian codebooks, and holds for any linear processing scheme and any pilot-based channel estimation scheme.
Since the bound is in terms of integrals that are not known in closed form and need to be evaluated numerically, which is impractical when the targeted error probability is low, we also present an accurate and easy-to-compute approximation, based on the saddlepoint method~\cite[Ch. XVI]{feller71-a}.

We then use the bound to evaluate the error probability in the \gls{ul} and \gls{dl} of a Massive MIMO network, with imperfect channel state information, pilot contamination, and spatially correlated channels. 
Both \gls{mmse} and \gls{mr} processing are considered. 
We remark that the application of the RCUs bound and saddlepoint approximation to characterize the error probability in this scenario is novel.
Furthermore, differently from~\cite{lancho20_04}, the proposed saddlepoint approximation involves quantities that can be characterized in closed form.
Hence, it can be evaluated efficiently.
We prove that the average error probability at finite blocklength with \gls{mmse} tends to zero as $M\to \infty$, whereas it converges to a positive number when \gls{mr}  is used. 
These results are similar in flavor to those about Massive MIMO ergodic rates in the infinite-blocklength regime (see, e.g.,~\cite{Sanguinetti18} and~\cite{Sanguinetti20}). 

Through numerical experiments, we estimate the error probability achievable for finite values of $M$ and quantify the impact of spatial correlation and pilot contamination. 
Inspired by~\cite{haenggi16-04a}, we use the \textit{network availability} as performance metric, which we define as the fraction of \gls{ue} placements for which the per-link error probability, averaged over the small-scale fading and the additive noise, is below a given target.
In the asymptotic outage setting, this quantity is obtained by characterizing the metadistribution of the \gls{sir}~\cite{haenggi16-04a}.
At finite blocklength, the network availability turns out to be related to the metadistribution of the so called \textit{generalized information density}~\cite[Eq.~(3)]{martinez11-02a}.

The numerical experiments show that, for finite values of $M$, it is important to take into account spatial correlation to obtain realistic estimates of the error probability.
Furthermore, pilot contamination turns out to have a strong impact on performance.
Consider for example a network with four $75 \text{\! m}\times75 \text{\! m}$ cells, $K=10$ UEs, $M=100$ \gls{bs} antennas. 
Furthermore, assume a transmit power of $10\dBm$ in UL and DL, an error probability target of $10^{-5}$ and a fixed frame of $300$ symbols, which accommodates pilots and data transmission in UL and DL. 
Assume also that in each data transmission phase, $160$ information bits need to be conveyed with an error probability target of $10^{-5}$. 
For this scenario, a network availability above $90\%$ can be achieved with MMSE processing in UL and DL only if pilot contamination is avoided by allocating as many pilot symbols as the total number of \glspl{ue} in the network.
In contrast, when all cells use the same pilot sequences, a network availability just above $50\%$ is achieved despite the fact that the shorter duration of the pilot sequences allows for a larger number of channel uses in the data phase.
With \gls{mr} processing, the network availability remains below $50\%$ for both \gls{ul} and \gls{dl}, even when pilot contamination is avoided. These numerical results suggest the following guidelines for the design of Massive MIMO for \gls{urllc} applications: $i$) Pilot contamination must be avoided;
    $ii$) In line with~\cite{Sanguinetti20}, \gls{mmse} should be chosen in place of the simpler \gls{mr}.

\subsection{Paper Outline and Notation}
In Section~\ref{sec:fbl-intro}, we present the finite-blocklength framework that will be used to analyze and design Massive MIMO networks.
In Section~\ref{sec:simo}, the finite-blocklength framework is used to analyze the impact on the error probability of pilot contamination, spatial correlation, and of the number of BS antennas, by focusing on a single-cell network with two \glspl{ue}.
The analysis is extended to a general multicell multiuser setting in Section~\ref{sec:mimo}.
Some conclusions are drawn in Section~\ref{sec:conclusions}.

Lower-case bold letters are used for vectors and upper-case bold letters are used for matrices.
The circularly-symmetric Gaussian distribution is denoted by $\jpg(0,\sigma^2)$, where $\sigma^2$ denotes the variance.
We use $\Ex{}{\cdot}$ to indicate the expectation operator, and $\prob{\cdot}$ for the probability of a set. The natural logarithm is denoted by $\log(\cdot)$, and $Q(\cdot)$ stands for the Gaussian $Q$-function. The Frobenius and spectral norms of a matrix ${\bf X}$ are denoted by $\| {\bf X} \|_F$ and $\| {\bf X} \|_2$, respectively.
The operators $(\cdot)^{\Ttran}$, $(\cdot)^*$, and $(\cdot)^{\Htran}$ denote transpose, complex conjugate, and Hermitian transpose, respectively.
Finally, we use $\stackrel{d}{=}$ to denote equality in distribution while, for two random sequences $a_n$, $b_n$, we write $a_n \asymp b_n$ to indicate that $\lim_{{n\to \infty}} (a_n -b_n) = 0$ almost surely.
\subsection{Reproducible Research}
The Matlab code used to obtain the simulation results is available at: \url{https://github.com/infotheorychalmers/URLLC_Massive_MIMO}.

\section{A Finite-Blocklength Upper-Bound on the Error Probability}\label{sec:fbl-intro}
In this section, we present a finite-blocklength upper bound on the error probability and  describe an efficient method for its numerical evaluation,  based on the saddlepoint approximation \cite[Ch. XVI]{feller71-a}.
We start by considering the simple case in which the received signal is the superposition of a scaled version of the desired signal and additive Gaussian noise. 
This simple channel model constitutes the building block for the analysis of the error probability achievable in the Massive MIMO networks considered in Sections~\ref{sec:simo} and~\ref{sec:mimo}.
\subsection{Upper Bound for Deterministic and Random Channels}
Consider a discrete \gls{awgn} channel given by 
\begin{equation}\label{eq:simplified_channel}
     v[k] = g q[k] + z[k], \quad k=1,\dots,n
\end{equation}
where $q[k]\in\mathbb{C}$ and $v[k]\in\mathbb{C}$ are the input and output over channel use $k$, respectively, and $n$ is the codeword length. 
Furthermore, $g\in\mathbb{C}$ is the channel gain, which is assumed to remain constant during transmission of the $n$-length codeword. The additive noise variables $\{z[k]\in\mathbb{C}; k=1,\ldots,n\}$, are \gls{iid}, $\jpg(0,\sigma^2)$, random variables. 
In what follows, we assume that:
  \begin{enumerate}
    \item The receiver \emph{does not know} the channel gain $g$ but has an estimate $\widehat{g}$ of $g$ that is treated as perfect.
    \item To determine the transmitted codeword $\vecq=[q[1],\dots,q[n]]^{\Ttran}$, the receiver seeks the codeword $\widetilde \vecq$ from the codebook $\mathcal {C}$ that, once scaled by 
    $\widehat{g}$, is the closest to the received vector $\vecv=[v[1],\dots,v[n]]^{\Ttran}\in\mathbb{C}^n$ in Euclidean distance.
    Mathematically, the estimated codeword $\widehat \vecq$ is obtained as
    \begin{equation}\label{eq:mismatched_snn_decoder}
      \widehat \vecq=\argmin_{\widetilde \vecq\in\mathcal{C}} \vecnorm{\vecv-\widehat{g}\widetilde \vecq}^2.
    \end{equation}
     A receiver operating according to~\eqref{eq:mismatched_snn_decoder} is known as mismatched \gls{snn} decoder~\cite{lapidoth02-05a}.
  Note that it coincides with the optimal maximum likelihood decoder \emph{if and only if} $\widehat{g}=g$.
  \end{enumerate}

We are interested in deriving an upper bound on the error probability $\epsilon=\prob{\widehat \vecq\neq \vecq}$ achieved by the \gls{snn} decoding rule \eqref{eq:mismatched_snn_decoder}. To do so, we follow a standard practice in information theory and use a random-coding approach \cite{gallager68a}. 
Specifically, we consider a Gaussian random code ensemble, where the elements of each codeword are drawn independently from a $\jpg(0,\rho)$ distribution.\footnote{Note that this ensemble is not optimal at finite blocklength, not even if $\widehat{g}=g$.
However, it is commonly used to obtain tractable expressions and insights into the performance of communication systems~\cite{marzetta16-a,bjornson19,MolavianJazi14}. Our analysis can be extended to other ensembles---see, e.g.,~\cite{ostman19-02}.}
Here, $\rho$ can be thought of as the average transmit power. 
We consider the cases where the channel gain $g$ in~\eqref{eq:simplified_channel} can be modelled as a deterministic or a random variable. In the literature, this latter case is commonly referred to as quasi-static fading setting~\cite[p. 2631]{biglieri98-10a}.

\begin{theorem}\label{thm:rcus}
    Assume that $g\in\mathbb{C}$ and $\widehat{g}\in\mathbb{C}$ in~\eqref{eq:simplified_channel} are deterministic.  There exists a coding scheme with $m= 2^b$ codewords of length $n$ operating according to the mismatched \gls{snn} decoding rule~\eqref{eq:mismatched_snn_decoder}, whose error probability $\epsilon$ is upper-bounded by\footnote{Note that the probability in~\eqref{eq:rcus_tail} is computed with respect to the channel inputs $\{q[k]\}_{k=1}^n$, the additive noise $\{z[k]\}_{k=1}^n$, and the random variable $u$.}
    \begin{IEEEeqnarray}{lCl}
      \epsilon &=& \prob{\widehat \vecq\neq \vecq}\nonumber\\
       &\leq& \prob{\sum_{k=1}^n {\imath_s(q[k],v[k])} + \log \lro{ u }\leq \log(m-1)} \label{eq:rcus_tail}
   \end{IEEEeqnarray}
   for all $s>0$. 
   Here, $u$ is a random variable that is uniformly distributed over the interval $[0,1]$ and $\imath_s(q[k],v[k])$ is the generalized information density, given by
   \begin{multline}
    \imath_s(q[k],v[k]) = -s \left|{v[k] - \widehat{g} q[k]}\right|^2 \\ + \frac{s\abs{v[k]}^2}{1+s\rho\abs{\widehat{g}}^2} + \log\lro{1+s\rho\abs{\widehat{g}}^2}.
  \label{eq:simple_infodens}
  \end{multline}
  Assume now that $g\in\mathbb{C}$ and $\widehat{g}\in\mathbb{C}$ in~\eqref{eq:simplified_channel} are random variables drawn according to an arbitrary joint distribution. 
  Then, for all $s>0$, the error probability $\epsilon$ is upper-bounded by%
\begin{IEEEeqnarray}{lCl}
  \epsilon &=& \prob{\widehat \vecq\neq \vecq}\nonumber\\
   &\leq& \Ex{g,\widehat{g}}{\prob{\sum_{k=1}^n {\imath_s(q[k],v[k])} \leq \log\frac{m-1}{u} \bigg\given g, \widehat{g}}} \label{eq:rcus_fading}
\end{IEEEeqnarray}
where the average is taken over the joint distribution of $g$ and $\widehat{g}$. 
If $g\in\mathbb{C}$ is a random variable and $\widehat{g}\in\mathbb{C}$ is deterministic,\footnote{This case will turn out important to analyze the \gls{dl} of Massive MIMO networks.} the average in \eqref{eq:rcus_fading} is only taken over the distribution of $g$.
\end{theorem}

\begin{IEEEproof}
  The proof for the case of $g$ and $\widehat{g}$ being deterministic, which is given in Appendix~A for completeness, follows by particularizing the \gls{rcus} bound introduced in~\cite[Th.~1]{martinez11-02a} to the considered setup. 
  The upper bound for random $g$ and $\widehat{g}$ readily follows by taking an expectation over the joint distribution of $g$ and $\widehat{g}$.
\end{IEEEproof}
Coarsely speaking, Theorem~\ref{thm:rcus} shows that the error probability in the finite-blocklength regime can be characterized in terms of the probability that the empirical average of the generalized information density $\imath_s$ is smaller than the chosen rate $R=(\log m)/n$.
In contrast, in the infinite-blocklength regime, the error (outage) probability, is given by the probability that the so-called generalized mutual information~\cite[Sec.~III]{lapidoth02-05a}  $I_s=\Ex{}{\imath_s(q[1],v[1])}$ is below the chosen rate. 
If $g$ is known at the receiver, i.e., $\widehat{g}=g$, it follows immediately from the decoding rule~\eqref{eq:mismatched_snn_decoder} that $\epsilon\to 0$ when the SNR grows unboundedly, i.e., $\rho/\sigma^2\to \infty$.
The following lemma shows that this is also true for the upper bounds~\eqref{eq:rcus_tail} and~\eqref{eq:rcus_fading}. 
\begin{lemma}\label{lem:inf-snr}
  If $g = \widehat g$, then
  \begin{IEEEeqnarray}{rCL}\label{eq:rcus_noisefree}
    \lim_{\rho/\sigma^2\to \infty} \prob{\sum_{k=1}^n {\imath_s(q[k],v[k])} \leq \log\frac{m-1}{u}} = 0.
  \end{IEEEeqnarray}
\end{lemma}

\begin{IEEEproof}
  This result is easily established by setting $v[k]=gq[k]$ and $\widehat{g}=g$ in~\eqref{eq:simple_infodens} and by noting that one can make~\eqref{eq:simple_infodens} arbitrarily large by choosing $s$ sufficiently large.
\end{IEEEproof}
 We anticipate that Lemma~\ref{lem:inf-snr} will be important for the characterization of the error probability of Massive MIMO in the asymptotic limit of large antenna arrays, i.e., $M\to \infty$.

The upper bounds in~\eqref{eq:rcus_tail} and~\eqref{eq:rcus_fading} involve the evaluation of a tail probability, which is not known in closed form and needs to be evaluated numerically. 
Furthermore, they can be tightened by performing an optimization over the parameter $s > 0$, which also needs to be performed numerically. 
All this is computational demanding, especially when one targets the low error probabilities required in \gls{urllc} applications. 
In the next section, we discuss how this problem can be alleviated by using a saddlepoint approximation. 
\subsection{Saddlepoint Approximation}\label{sec:saddlepoint-approximation}
One possible way to numerically approximate~\eqref{eq:rcus_tail} and~\eqref{eq:rcus_fading} is to perform a normal approximation on the probability term based on the Berry-Esseen central limit theorem~\cite[Ch.~XVI.5]{feller71-a}.
This leads to the following expansion:
\begin{multline}\label{eq:normal_approximation}
  \prob{\sum_{k=1}^n {\imath_s(q[k],v[k])} \leq \log\frac{m-1}{u}} \\ = Q\lefto(\frac{nI_s-\log(m-1)}{\sqrt{nV_s}}\right) + \landauo\lefto(\frac{1}{\sqrt{n}}\right)
\end{multline}
where $I_s=\Ex{}{\imath_s(q[1],v[1])}$ is the so-called generalized mutual information~\cite[Sec.~III]{lapidoth02-05a},
\begin{equation}
  V_s=\Ex{}{\abs{\imath_s(q[1],v[1])-I_s}^2}
\end{equation}
is the variance of the information density, typically referred to as \emph{channel dispersion}~\cite[Sec.~IV]{polyanskiy10-05a}, and $\landauo\lefto({1}/{\sqrt{n}}\right)$ accounts for terms that decay faster than $1/\sqrt{n}$ as $n\to\infty$.
The so-called \emph{normal approximation} obtained by neglecting the $\landauo\lefto({1}/{\sqrt{n}}\right)$ term in~\eqref{eq:normal_approximation} is accurate only when $R=(\log m)/n$ is close to $I_s$~\cite{lancho20_04}. 
Unfortunately, this is typically not the case in \gls{urllc} since one needs to operate at rates much lower than $I_s$ to obtain the required low error probabilities at SNR values of practical interest (see, e.g., \cite[Fig.~3]{lancho20_04}). 
A more accurate approximation, that holds for all values of~$R$, can be obtained using the saddlepoint method.
The main idea of the saddlepoint method is to perform an exponential tilting~\cite[Ch.~XVI.7]{feller71-a} on the random variables $\{\imath_s(q[k],v[k]), k =1,\ldots,n\}$, which moves their mean close to the desired rate~$R$. This guarantees that a subsequent use of the normal approximation yields small errors.

The saddlepoint method has been applied to obtain accurate approximations of the \gls{rcus} in, e.g., \cite{scarlett14-05a}  and~\cite{lancho20_04}. 
In the following, we particularize these expressions to the setup considered in Theorem~\ref{thm:rcus} and refer to~\cite{scarlett14-05a,lancho20_04} for further details and proofs. 
While to obtain~\eqref{eq:normal_approximation}, it is sufficient to check that the third central moment of $\imath_s(q[k],v[k])$ is bounded (which is indeed the case in our setup), the existence of a saddlepoint approximation requires the more stringent condition that the third derivative of the \gls{mgf} of $-\imath_s(q[k],v[k])$ exists in a neighborhood of zero.
Specifically, we require that there exist two values $\underline{\zeta} < 0 < \overline{\zeta}$
such that
\begin{equation}
  \sup_{\underline{\zeta} < \zeta < \overline{\zeta}} \frac{d^3}{d \zeta^3} \Bigl|\Ex{}{e^{-\zeta \imath_s(q[k],v[k])}} \Bigr| < \infty.\label{eq:cond_saddle}
\end{equation}
As shown in Appendix~B, this condition is verified in our setup.
Specifically, we have that 
\begin{IEEEeqnarray}{lCl}
  \underline{\zeta} &=& -\frac{\sqrt{(\beta_B-\beta_A)^2 + 4\beta_A\beta_B(1-\nu)}+\beta_A-\beta_B}{2\beta_A\beta_B(1-\nu)}\label{eq:RoC_values_A}\IEEEeqnarraynumspace\\
  \overline{\zeta} &=& \frac{\sqrt{(\beta_B-\beta_A)^2 + 4\beta_A\beta_B(1-\nu)} -\beta_A+\beta_B}{2\beta_A\beta_B(1-\nu)}\label{eq:RoC_values_B}\IEEEeqnarraynumspace
\end{IEEEeqnarray}
where
\begin{IEEEeqnarray}{lCl}
  \beta_A &=& s(\rho \abs{g-\widehat{g}}^2+\sigma^2)\label{eq:betaA}\\
  \beta_B &=& \frac{s}{1+s\rho\abs{\widehat{g}}^2} \lro{\rho\abs{g}^2 + \sigma^2}\label{eq:betaB}\\
  \nu 
  &=&\frac{s^2 \left|{\rho  \abs{g}^2 + \sigma^2-g^*\widehat{g}\snr}\right|^2}{\beta_A \beta_B (1+s \rho \abs{\widehat{g}}^2)}.
  \label{eq:corr_coeff_SISO}
\end{IEEEeqnarray}

The saddlepoint approximation that will be provided in Theorem~\ref{thm:saddlepoint} below depends on the \gls{cgf} of $-\imath_s(q[k],v[k])$
\begin{IEEEeqnarray}{lCl}
  \kappa(\zeta) &=& \log \Ex{}{e^{-\zeta \imath_s(q[k],v[k])}}
    \label{eq:cgf_def}
\end{IEEEeqnarray}
and on its first derivative $\kappa'(\zeta)$ and second derivative $\kappa''(\zeta)$. 
In our setup, these quantities can be computed in closed form for all $\zeta \in (\underline{\zeta},\overline{\zeta})$ and are given by (see Appendix B)
\begin{IEEEeqnarray}{lCl}
  \kappa(\zeta) 
     &=&{}
    -\zeta\log\lro{1+s\rho\abs{\widehat{g}}^2}\nonumber\\
    &&{} - \log \lro{1+\lro{\beta_B-\beta_A}\zeta -\beta_A\beta_B(1-\nu)\zeta^2}\nonumber\\\label{eq:cgf}\\
    \kappa'(\zeta) &=&{}
    -\log\lro{1+s\rho\abs{\widehat{g}}^2} \nonumber\\
    &&{} - \frac{\lro{\beta_B-\beta_A} -2\beta_A\beta_B(1-\nu)\zeta}{1+\lro{\beta_B-\beta_A}\zeta -\beta_A\beta_B(1-\nu)\zeta^2} \label{eq:cgf_1} \\
    \kappa''(\zeta) &=& \lrho{\frac{\lro{\beta_B-\beta_A} -2\beta_A\beta_B(1-\nu)\zeta}{1+\lro{\beta_B-\beta_A}\zeta -\beta_A\beta_B(1-\nu)\zeta^2}}^2 \nonumber\\
    &&{} +  \frac{2\beta_A\beta_B(1-\nu)}{1+\lro{\beta_B-\beta_A}\zeta -\beta_A\beta_B(1-\nu)\zeta^2}. \label{eq:cgf_2}
\end{IEEEeqnarray} 
Note that $-\kappa(\zeta)$ coincides with the so-called Gallager's $E_0$ function for the mismatched case~\cite[Eq.~(22)]{martinez11-02a}.
As a consequence, we have that $I_s=-\kappa'(0)$. 
Furthermore, the so-called \textit{critical rate} $R_s^{\mathrm{cr}}$ (see~\cite[Eq. (5.6.30)]{gallager68a}) is given by
\begin{equation}
  R_s^{\mathrm{cr}} = -\kappa'(1).
\end{equation}
We are now ready to present the saddlepoint expansion of the \gls{rcus} bound~\eqref{eq:rcus_tail}.
\begin{theorem}\label{thm:saddlepoint}
  Let $m=e^{nR}$ for some $R>0$, and let $\zeta\in(\underline{\zeta},\overline{\zeta})$ be the solution to the equation $R=-\kappa'(\zeta)$.\footnote{The existence of such a solution for all rates $R\geq0$ follows from~\eqref{eq:cgf_1}.} 
  If $\zeta\in[0,1]$, then $R_s^{\mathrm{cr}} \leq R \leq I_s$ and
  \begin{IEEEeqnarray}{lCl}
    \IEEEeqnarraymulticol{3}{l}{
    \prob{\sum_{k=1}^n {\imath_s(q[k],v[k])} \leq \log\frac{e^{nR}-1}{u}}}\nonumber\\* 
    & = & e^{n[\kappa(\zeta)+\zeta R]}\lrho{\Qexp\lro{\zeta}+\Qexp\lro{1-\zeta}+o\lro{\frac{1}{\sqrt{n}}}} \label{eq:saddlepoint_U_pos}\IEEEeqnarraynumspace
  \end{IEEEeqnarray}
  where 
  \begin{IEEEeqnarray}{rCl}
    \Qexp\lro{u} & \triangleq & e^{n\frac{u^2}{2}\kappa''(\zeta)}Q\lro{u\sqrt{n\kappa''(\zeta)}}\IEEEeqnarraynumspace\label{eq:help_fcn_theta}
  \end{IEEEeqnarray}
  and $o(1/\sqrt{n})$ comprises terms that vanish faster than $1/\sqrt{n}$ and are uniform in $\zeta$.

  If $\zeta>1$, then $R < R_s^{\mathrm{cr}}$ and
  \begin{IEEEeqnarray}{lCl}
    \IEEEeqnarraymulticol{3}{l}{
    \prob{\sum_{k=1}^n {\imath_s(q[k],v[k])} \leq \log\frac{e^{nR}-1}{u}}}\nonumber\\* 
       &=& e^{n[\kappa(1)+ R]}\lrho{\widetilde{\Psi}_n(1,1)+\widetilde{\Psi}_n(0,-1)+\mathcal{O}\lro{\frac{1}{\sqrt{n}}}} \label{eq:saddlepoint_U_pos_cr}\IEEEeqnarraynumspace
  \end{IEEEeqnarray}
  where 
  \begin{IEEEeqnarray}{lCl}
    \widetilde{\Psi}_n(a_1,a_2) & = & e^{na_1\lrho{R_s^{\mathrm{cr}}-R+\frac{\kappa''(1)}{2}}}\nonumber\\
    &&\times{} Q\lro{a_1\sqrt{n\kappa''(1)}+a_2\frac{n(R_s^{\mathrm{cr}}-R)}{\sqrt{n\kappa''(1)}}}\label{eq:help_fcn_theta_2}\IEEEeqnarraynumspace
  \end{IEEEeqnarray}
  and $\mathcal{O}(1/\sqrt{n})$ comprises terms that are of order $1/\sqrt{n}$ and are uniform in $\zeta$. 
  If $\zeta<0$, then $R > I_s$ and
  \begin{IEEEeqnarray}{lCl}
    \IEEEeqnarraymulticol{3}{l}{
    \prob{\sum_{k=1}^n {\imath_s(q[k],v[k])} \leq \log\frac{e^{nR}-1}{u}}}\nonumber\\* \qquad
    &=& 1 - e^{n[\kappa(\zeta)+\zeta R]}\biggl[ \Qexp(-\zeta) - \Qexp\lro{1-\zeta} \nonumber\\
     &&\qquad\qquad\qquad\quad {} +o\lro{\frac{1}{\sqrt{n}}}\biggr].\label{eq:saddlepoint_U_neg}
  \end{IEEEeqnarray}
\end{theorem}
\begin{IEEEproof}
The proof follows along steps similar to~\cite[App. E]{scarlett14-05a} and to \cite[App. I]{lancho20_04}, and it is thus omitted because of space limitations. 
\end{IEEEproof}

We will refer to the approximations obtained by ignoring the $o(1/\sqrt{n})$ terms and the $\mathcal{O}(1/\sqrt{n})$ terms in~\eqref{eq:saddlepoint_U_pos}, \eqref{eq:saddlepoint_U_pos_cr}, and~\eqref{eq:saddlepoint_U_neg} as \emph{saddlepoint approximations}.
Note that the exponential term on the right-hand side of~\eqref{eq:saddlepoint_U_pos} and~\eqref{eq:saddlepoint_U_pos_cr} corresponds to the Gallager error exponent for the mismatch decoding scenario~\cite{Kaplan93}.
This means that the saddlepoint approximation provides an estimate of the subexponential factor, thereby allowing one to obtain accurate approximations of error probability values for which the error exponent is inaccurate.  
In a nutshell,  the key  idea  of  the  saddlepoint  method  is  to  isolate the Gallager error-exponent term, i.e., the exponential term in~\eqref{eq:saddlepoint_U_pos},~\eqref{eq:saddlepoint_U_pos_cr}, and~\eqref{eq:saddlepoint_U_neg}, which governs the exponential decay of the error probability as a function of the blocklength, and then to use the Berry-Esseen central-limit theorem to characterize only the pre-exponential factor, i.e., the factor that multiplies the exponential term.
It is also worth highlighting that since all quantities in~\eqref{eq:saddlepoint_U_pos},~\eqref{eq:saddlepoint_U_pos_cr}, and~\eqref{eq:saddlepoint_U_neg} are known is closed form, the evaluation of the saddlepoint approximation, for a given $\zeta$ and its corresponding rate $R=-\kappa'(\zeta)$, entails a complexity similar to that of the normal approximation~\eqref{eq:normal_approximation}.

Note that both the saddlepoint approximation and the normal approximation can be tightened by performing an optimization over $s$, which may be time consuming. 
One way to avoid this step is to choose an $s$ that is optimal in some asymptotic regime.
One can for example set $s$ so as to maximize the generalized mutual information $I_s$.
The corresponding value for $s$ can be obtained in closed form~\cite[Eq.~(64)]{lapidoth02-05a}.
\subsection{Outage Probability and Normal Approximation}
Equipped with the bound~\eqref{eq:rcus_fading} and with an efficient method for the numerical evaluation of the probability term within~\eqref{eq:rcus_fading}, we can now evaluate the error probability achievable for short blocklengths and investigate whether the outage probability is an accurate performance metric in Massive MIMO systems for \gls{urllc} applications.
For the sake of simplicity, we consider a single-UE multiantenna system in which the \gls{bs} has a large number $M$ of antennas. 
We denote by ${\bf h}\in \mathbb{C}^M$ the channel between the UE and the BS array and assume that it can be modelled as uncorrelated Rayleigh fading ${\bf h}\sim \jpg({\bf 0}_M, \beta{\bf I}_M)$ where $\beta$ is the large-scale fading gain~\cite[Sec. 1.3.2]{bjornson19}. 
If perfect \gls{csi} is available at the receiver and \gls{mr} combining is used for detection, the \gls{ul} channel input-output relation can be expressed as 
\begin{equation}\label{eq:simplified_channel_fading}
     v[k] = \frac{{\bf h}^{\Htran}}{\vecnorm{\bf h}} {\bf h} q[k] + \frac{{\bf h}^{\Htran}}{\vecnorm{\bf h}} {\bf z}^\prime[k], \quad k=1,\dots,n
\end{equation}
where ${\bf z}^\prime[k] \sim \jpg({\bf 0}_M, \sigma^2{\bf I}_M)$ is the thermal noise over the antenna array over channel use $k$. 
Note that \eqref{eq:simplified_channel_fading} can be mapped into~\eqref{eq:simplified_channel} by setting $g = \frac{{\bf h}^{\Htran}}{\vecnorm{\bf h}} {\bf h}=\vecnorm{\bf h}$  and $z[k] = \frac{{\bf h}^{\Htran}}{\vecnorm{\bf h}} {\bf z}^\prime[k]\sim \jpg({0}, \sigma^2)$. Since ${\bf h}$ is perfectly known at the receiver, we have that $\widehat{g}=g=\vecnorm{\bf h}$.
In the limit $n\to\infty$, it can be shown that the probability term in~\eqref{eq:rcus_fading}, once optimized over the parameter $s$, is equal to $1$ if $\log(1+\snr\abs{g}^2/\sigma^2)<R$ and $0$ otherwise.
This means that the bound in~\eqref{eq:rcus_fading} converges to the outage probability
\begin{equation}
  \prob{\log\left(1+\frac{\snr{g}^2}{\sigma^2}\right)<R}\label{eq:outage}.
\end{equation}
Here, the probability is evaluated with respect to the random variable $g = \vecnorm{{\bf h}}$. 
\begin{figure}[t]     
  \centering
  \begin{subfigure}{0.45\textwidth}
    \centering
    \includegraphics[width=\textwidth]{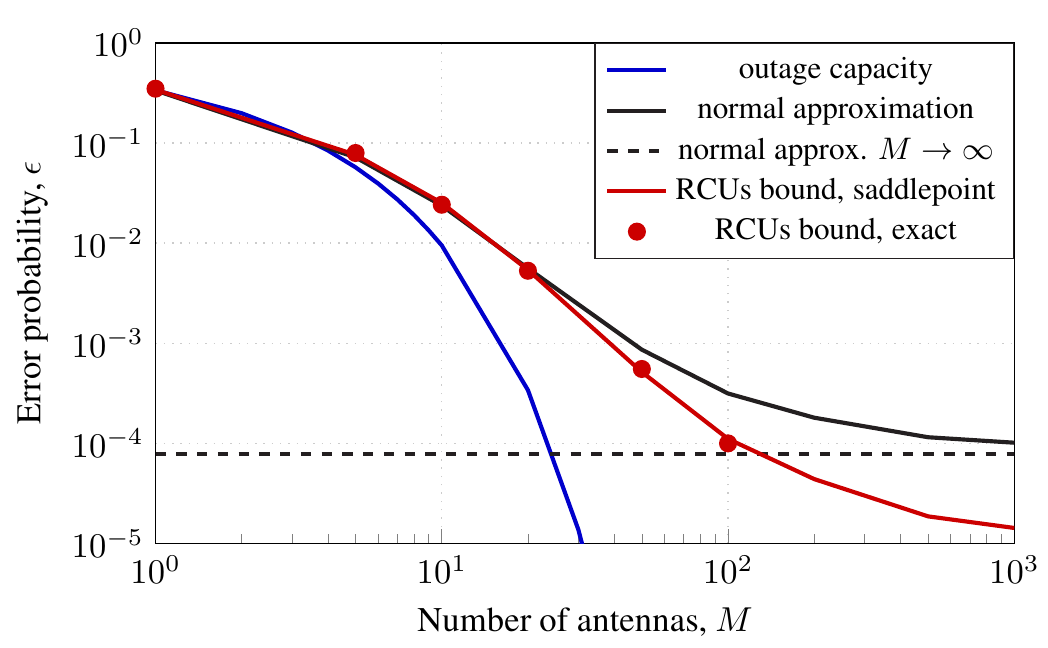}
    \caption{Fixed average received $\text{SNR} = 1 \dB$.}
    \label{fig:secIIa}
  \end{subfigure}
  \hspace{3mm}
  \begin{subfigure}{0.45\textwidth}
    \centering
    \includegraphics[width=\textwidth]{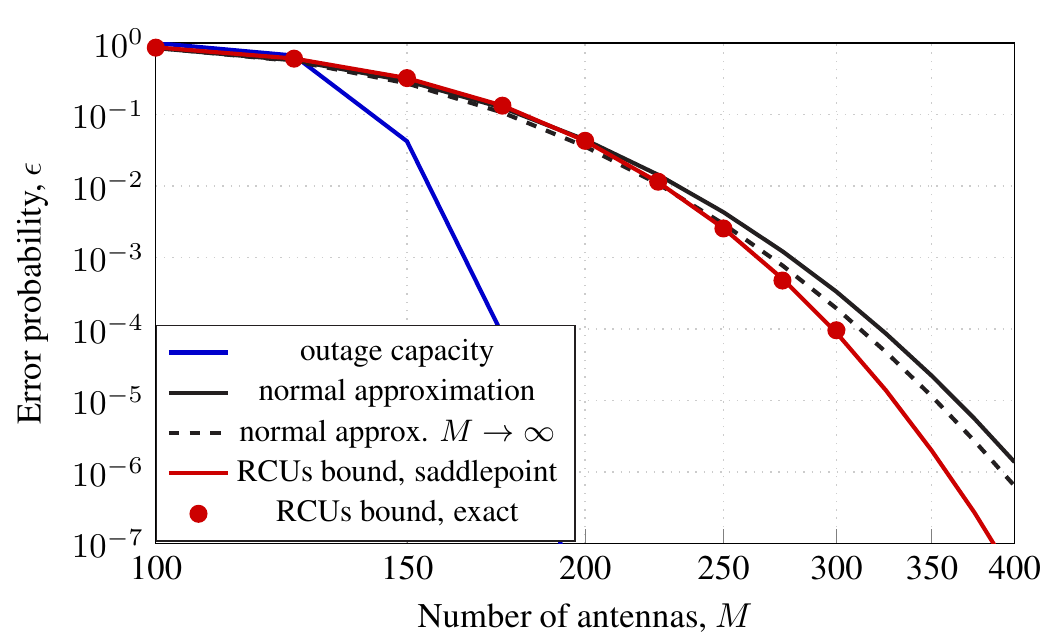}
    \caption{Fixed transmit power $\rho = -24 \dBm$.}
    \label{fig:secIIb}
  \end{subfigure}
  \caption{Average error probability in the \gls{ul} of a single-\gls{ue} multiantenna system when $\widehat{g}=g = \vecnorm{\bf h}$ with ${\bf h}\sim \jpg({\bf 0}_M, \beta{\bf I}_M)$, $n=100$, and $R=0.6$ bits per channel use.
  The \gls{ue} is assumed to be at a distance from the \gls{bs} that results in $\beta/\sigma^2 = 1$.
     }
       \label{fig:secII}
  \end{figure}

In Fig.~\ref{fig:secII}, we depict the outage probability in~\eqref{eq:outage} as a function of the number of \gls{bs} antennas~$M$. Comparisons are made with the upper bound in~\eqref{eq:rcus_fading}, evaluated by means of both Monte-Carlo integration (exact) and the saddlepoint approximation in Theorem~\ref{thm:saddlepoint}. 
We also depict the normal approximation obtained by averaging~\eqref{eq:normal_approximation} over $g$.
In the evaluation of~\eqref{eq:rcus_fading}, we set $\widehat{g}=g$ and optimize over the parameter $s$ by means of a bisection search.\footnote{In all numerical simulations presented throughout the paper, we will always evaluate the error probability bound in~\eqref{eq:rcus_fading} using the saddlepoint approximation in Theorem~\ref{thm:saddlepoint}, and optimize it over the parameter $s$ via a bisection search.}
We assume that $\sigma^2 = -94\dBm$ and set $\beta=\sigma^2$ so that $\Ex{}{{g}^2}/\sigma^2 =\beta M/\sigma^2= M$.\footnote{With the distance-dependent pathloss model that will be introduced in~\eqref{eq:pathloss-coefficient}, this corresponds to a distance of $36.4$\,m.}
Furthermore, we consider a codeword length $n=100$ and a rate of $R=60/100=0.6$ bits per channel use.  
   
In Fig.~\ref{fig:secIIa}, we illustrate the error probability for a transmit power $\rho$ that decreases as $1/M$. 
Specifically, we set $\rho=\widetilde\rho/M$ with $\widetilde\rho = 1$\,dB.   
Since ${g}^2/M\to \beta$ as $M\to\infty$ and we assume $\beta =\sigma^2$, it thus follows that the instantaneous SNR $ \rho {g}^2/\sigma^2$ converges to the deterministic value $\widetilde \rho$ as $M\to \infty$. 
This means that, as $M\to \infty$, the normal approximation for \iid Gaussian inputs given in~\eqref{eq:normal_approximation} for a fixed $g$ converges to a deterministic quantity. 
Specifically, in the limit $M\to\infty$, we have that  
$I_s=\log(1+ \widetilde\rho\beta/\sigma^2)$ (achieved for $s=1/\sigma^2$) and $V_s = 2 \widetilde\rho\beta/( \widetilde\rho\beta+\sigma^2)$~\cite[Eq. (2.55)]{MolavianJazi14}.
The resulting approximation is of interest because it does not require any Monte-Carlo averaging over the realizations of the fading channel. 
From  Fig.~\ref{fig:secIIa}, we see that the outage probability~\eqref{eq:outage} approximates well the exact RCUs bound~\eqref{eq:rcus_fading} only when $M$ is small, i.e., $M<5$, whereas the normal approximation loses accuracy when $M>20$.
Both approximations are not accurate at the low error probabilities of interest in URLLC.
The saddlepoint approximation is instead very accurate for all $M$ values.

In Fig.~\ref{fig:secIIb} we report the error probability with no power scaling so that the average received SNR increases as $M$ increases.  
Specifically, we consider a fixed transmit power $\rho=-24\dBm$.
Hence, for $M=320$ the average received SNR in Fig.~\ref{fig:secIIb} equals $1\dB$, which coincides with the average received SNR in Fig.~\ref{fig:secIIa}.
With no power scaling, the outage probability~\eqref{eq:outage} is an accurate approximation for the RCUs bound~\eqref{eq:rcus_fading} only for very large values of the error probability, whereas the accuracy of the normal approximation~\eqref{eq:normal_approximation} is acceptable for $\epsilon$ within the range $[10^{-3},1]$.
The saddlepoint approximation again is on top of the \gls{rcus} bound for all values of error probability considered in the figure.

Based on the above results, we conclude that outage probability and the normal approximation do not always provide accurate estimates of the error probability achievable in large-antenna systems with short-packet communications over quasi-static channels. 
The accuracy of these approximations becomes even more questionable in the presence of imperfect \gls{csi}. 
This problem can be avoided altogether by using the nonasymptotic bound~\eqref{eq:rcus_fading} in Theorem~\ref{thm:rcus}, which can be efficiently evaluated by means of the saddlepoint approximation in Theorem~\ref{thm:saddlepoint}. 
In the next two sections, we will show how the simple input-output relation~\eqref{eq:simplified_channel} can be used as building block for the analysis of practical Massive MIMO networks with imperfect \gls{csi}, pilot contamination, spatial correlation among antennas, and both inter-cell and intra-cell interference. 
Theorem~\ref{thm:rcus} and Theorem~\ref{thm:saddlepoint} will then be used to efficiently evaluate the average error probability also in these more realistic scenarios.


\section{A Two-UE Single-Cell Massive MIMO Scenario}\label{sec:simo}
We consider a single-cell network where the \gls{bs} is equipped with $M$ antennas and serves $K=2$ single-antenna \glspl{ue}. 
We denote by $\vech_{i}\in \mathbb{C}^{M}$ the channel vector between the \gls{bs} and \gls{ue} $i$ for $i=1,2$. 
We use a correlated Rayleigh fading model where $\vech_i \sim \jpg({\bf 0}_M,{\bf R}_i)$ remains constant for the duration of a codeword transmission. 
The normalized trace $\beta_i = \tr({\bf{R}}_i)/M$ determines the average large-scale fading between UE $i$ and the BS, while the eigenstructure of ${\bf{R}}_i$ describes its spatial channel correlation~\cite[Sec. 2.2]{bjornson19}. 
We assume that ${\bf{R}}_1$ and ${\bf{R}}_2$   are known at the BS; see, e.g., \cite{sanguinetti16,Sanguinetti20} for a description of practical estimation methods. 
This setup is sufficient to demonstrate the usefulness of the framework developed in Section II for the analysis and design of Massive MIMO networks. 
A more general setup will be considered in Section IV.
\subsection{Uplink pilot transmission}\label{sec:pilots}
We consider the standard \gls{tdd} Massive MIMO protocol, where the \gls{ul} and \gls{dl} transmissions are assigned $n$ channel uses in total, divided in $\np$ channel uses for \gls{ul} pilots, $\nul$ channel uses for \gls{ul} data, and  $\ndl=n-\np-\nul$ channel uses for \gls{dl} data. 
We assume that the $\np$-length pilot sequence $\bphiu_i \in \mathbb{C}^{\np}$ with $\bphiu_i^{\Htran}\bphiu_i = \np$ is used by \gls{ue} $i$ for channel estimation. 
The elements of $\bphiu_i$ are scaled by the square-root of the pilot power $\sqrt{\rho^{\mathrm{ul}}}$ and transmitted over $\np$ channel uses. 
When the \glspl{ue} transmit their pilot sequences, the received pilot signal $ {\bf Y}^{\mathrm{pilot}}\in \mathbb{C}^{M\times\np}$ is
\begin{IEEEeqnarray}{lCl}
  {\bf Y}^{\mathrm{pilot}} = \sqrt{\rho^{\mathrm{ul}}}\vech_1 \bphiu_1^{\Htran} + \sqrt{\rho^{\mathrm{ul}}}\vech_2 \bphiu_2^{\Htran} +   {\bf Z}^{\mathrm{pilot}}  \label{eq:simo_channel_ul_pilot}
\end{IEEEeqnarray}
 where ${\bf Z}^{\mathrm{pilot}} \in \mathbb{C}^{M \times \np}$ is the additive noise with \gls{iid} elements distributed as $\jpg(0,\sigma_{\mathrm{ul}}^{2})$. 
 Assuming that $\mathbf{R}_1$ and $\mathbf{R}_2$ are known at the \glspl{bs}, the \gls{mmse} estimate of $\vech_i$ is \cite[Sec.~3.2]{bjornson19}
 \begin{align} \label{eq:MMSEestimator_h_1}
 \widehat{\vech}_i  = \sqrt{\rho^{\mathrm{ul}} \np}{\bf R}_i {\bf Q}_i ^{-1}  \left({\bf Y}^{\mathrm{pilot}}\bphiu_i\right)
 \end{align}
 for $i=1,2$ with 
  \begin{align}\label{eq:covariance-matrix-pilot-signal}
 {\bf Q}_i  = \rho^{\mathrm{ul}} {\bf R}_1\bphiu_1^{\Htran}\bphiu_i + \rho^{\mathrm{ul}} {\bf R}_2 \bphiu_2^{\Htran}\bphiu_i +  \sigma_{\mathrm{ul}}^2  {\bf I}_{M}.
  \end{align}
The \gls{mmse} estimate $ \widehat{\vech}_i $ and the estimation error $\widetilde{\vech}_i  = {\vech}_i  - \widehat{\vech}_i$ are independent random vectors, distributed as $\widehat{\vech}_i \sim \jpg ({\bf 0}, {\bf \Phi}_i )$ and $\widetilde{\vech}_i \sim \jpg ({\bf 0}, {\bf R}_i  - {\bf \Phi}_i )$, respectively, with ${\bf \Phi}_i  = {\rho^{\mathrm{ul}} \np}{\bf R}_i  {\bf Q}_i ^{-1} {\bf R}_i$.

It follows from~\eqref{eq:covariance-matrix-pilot-signal} that if the two \glspl{ue} use orthogonal pilot sequences, i.e., $\herm{\bphiu_1}\bphiu_2=0$, they do not interfere, whereas they interfere if they use the same pilot sequence, i.e. $\bphiu_1 = \bphiu_2$.
This interference is known as pilot contamination and has two main consequences in the channel estimation process \cite[Sec.~3.2.2]{bjornson19}. 
The first is a reduced estimation quality; the second is that the estimates $ \widehat{\vech}_1$ and  $\widehat{\vech}_2$ become correlated. 
To see this, observe that if $\bphiu_1 = \bphiu_2$ then ${\bf Y}^{\mathrm{pilot}}\bphiu_1 = {\bf Y}^{\mathrm{pilot}}\bphiu_2$ and $ {\bf Q}_1= {\bf Q}_2 = {\bf Q}$ with ${\bf Q}=  \rho^{\mathrm{ul}} \np{\bf R}_1 + \rho^{\mathrm{ul}}\np {\bf R}_2 +  \sigma_{\mathrm{ul}}^2  {\bf I}_{M}$. 
Hence, $  \widehat{\vech}_2$ can be written as $  \widehat{\vech}_2  = {\bf R}_2\left({\bf R}_1\right)^{-1}  \widehat{\vech}_1$
 provided that ${\bf R}_1$ is invertible.
This implies that the two estimates are correlated with cross-correlation matrix given by $\Ex{}{    \widehat{\vech}_1\widehat{\vech}_2^{\Htran}} ={\bf \Upsilon}_{12} = \rho^{\mathrm{ul}} \np{\bf R}_1 {\bf Q} ^{-1} {\bf R}_2$. 
This holds even though the underlying channels ${\bf h}_1$ and ${\bf h}_2$ are statistically independent, which implies that $\Ex{}{   {\vech}_1{\vech}_2^{\Htran}}  = {\bf 0}_M$. 
Observe that if there is no spatial correlation, i.e., ${\bf R}_i=\beta_i{\bf I}_M$, $i=1,2$, then the channel estimates are identical up to a scaling factor, i.e., they are linearly dependent.
We will return to the issue of pilot contamination in Section~\ref{sec:asymptotics}.
\subsection{Uplink data transmission}\label{sec:UL_2UES}
During \gls{ul} data transmission, the received complex baseband signal $\vecr^{\mathrm{ul}}[k]\in \mathbb{C}^{M}$ over an arbitrary channel use $k$, where $k=1,\ldots,\nul$, is given by
\begin{equation}
  \vecr^{\mathrm{ul}}[k]  = \vech_1 x_1^{\mathrm{ul}}[k] + \vech_2 x_2^{\mathrm{ul}}[k] + \vecz^{\mathrm{ul}}[k]\label{eq:simo_channel_ul_1}
\end{equation}
where $x_i^{\mathrm{ul}}[k] \sim \jpg (0, \rho^{\mathrm{ul}})$ is the information bearing signal\footnote{As detailed in Section~\ref{sec:fbl-intro}, we will evaluate the error probability for a Gaussian random code ensemble, where the elements of each codeword are drawn independently from a $\jpg (0, \rho^{\mathrm{ul}})$ distribution.} transmitted by \gls{ue} $i$ with $\rho^{\mathrm{ul}}$ being the average \gls{ul} transmit power and $\vecz^{\mathrm{ul}}[k] \sim \jpg ({\bf 0}, \sigma_{\mathrm{ul}}^{2} {\bf I}_{M})$ is the independent additive noise. 
The \gls{bs} detects the signal $x_1^{\mathrm{ul}}[k]$ by using the combining vector ${\bf u}_1 \in \mathbb{C}^{M}$, to obtain 
\begin{IEEEeqnarray}{lCl}
y_1^{\mathrm{ul}}[k] &=& \herm{{\bf u}}_1 \vecr^{\mathrm{ul}}[k]\nonumber\\
 &=& \herm{{\bf u}}_1\vech_1 x_1^{\mathrm{ul}}[k] + \herm{{\bf u}}_1\vech_2 x_2^{\mathrm{ul}}[k] + \herm{{\bf u}}_1 \vecz^{\mathrm{ul}}[k]. \label{eq:simo_channel_ul}
\end{IEEEeqnarray}
Note that~\eqref{eq:simo_channel_ul} has the same form as~\eqref{eq:simplified_channel} with $v[k] = y_1^{\mathrm{ul}}[k]$, $q[k]= x_1^{\mathrm{ul}}[k]$, $g=\herm{{\bf u}}_1 \vech_1$, and $z[k] =\herm{{\bf u}}_1\vech_2 x_2^{\mathrm{ul}}[k] + \herm{{\bf u}}_1 \vecz^{\mathrm{ul}}[k]$.
Furthermore, given $\{{\bf h}_1,{\bf u}_1,{\bf h}_2\}$, the random variables $\{z[k] : k=1,\ldots,\nul\}$ are conditionally \iid and $z[k]\sim \jpg(0, \sigma^2)$ with $\sigma^2 = \rho^{\mathrm{ul}}\abs{\herm{{\bf u}}_1\vech_2}^2 + \vecnorm{{\bf u}_1}^2 \sigma\sub{ul}^2$.

We assume that the \gls{bs} treats the acquired (noisy) channel estimate $\widehat{\vech}_1$ as perfect. 
This implies that, to recover the transmitted codeword, which we assume to be drawn from a codebook $\setC^\mathrm{ul}$, it performs mismatched \gls{snn} decoding with $\widehat{g} = \herm{{\bf u}}_1 \widehat{\vech}_1$. 
Specifically, the estimated codeword $\widehat{\bf x}_1^{\mathrm{ul}}$ is obtained as
\begin{equation}\label{eq:mismatched_snn_decoder-uplink}
  \widehat{\bf x}_1^{\mathrm{ul}}=\argmin_{\widetilde{\bf x}_1^{\mathrm{ul}} \in \setC^\mathrm{ul}} \vecnorm{{\bf y}_1^{\mathrm{ul}}-(\herm{{\bf u}}_1 \widehat{\vech}_1)\widetilde{\bf x}_1^{\mathrm{ul}}}^2
\end{equation}
with ${\bf y}_1^{\mathrm{ul}} = [y_1^{\mathrm{ul}}[1],\ldots,y_1^{\mathrm{ul}}[\nul]]^{\Ttran}$ and $\widetilde{\bf x}_1^{\mathrm{ul}} = [\widetilde{x}_1^{\mathrm{ul}}[1],\ldots, \widetilde{x}_1^{\mathrm{ul}}[\nul]]^{\Ttran}$. 
It thus follows that~\eqref{eq:rcus_tail} provides a bound on the conditional error probability for \gls{ue} $1$ given $g$ and $\widehat{g}$.
To obtain the average error probability, we need to take an expectation over $g=\herm{{\bf u}}_1 \vech_1$, $\widehat{g} = \herm{{\bf u}}_1 \widehat{\vech}_1$, and $\sigma^2=\rho^{\mathrm{ul}}\abs{\herm{{\bf u}}_1\vech_2}^2 + \vecnorm{{\bf u}_1}^2 \sigma\sub{ul}^2$, which results in
\begin{IEEEeqnarray}{lCl}
  &&\epsilon_1^{\mathrm{ul}}\nonumber \\ && \leq \Ex{}{\prob{\sum_{k=1}^{\nul} \imath_s(y_1^{\mathrm{ul}}[k],x_1^{\mathrm{ul}}[k]) \leq \log\frac{m-1}{u} \Bigg\given g, \widehat{g},\sigma^2}}.\nonumber\\ \label{eq:rcus_tail_simo_ul}
\end{IEEEeqnarray}
The saddlepoint approximation in Theorem~\ref{thm:saddlepoint} can be applied verbatim to efficiently compute the conditional probability in~\eqref{eq:rcus_tail_simo_ul}. 
The average error probability for \gls{ue}~$2$ can be evaluated similarly.
    
The combining vector ${\bf u}_1$ is selected at the \gls{bs} based on the channel estimates $\widehat{\bf h}_1$ and $\widehat{\bf h}_2$. 
The simplest choice is to use \gls{mr} combining: ${\bf u}_1^{\rm{MR}} =\widehat{\vech}_1/M$. 
A more computationally intensive choice is \gls{mmse} combining:
\begin{equation}\label{eq:MMSE_combiner}
{\bf u}_1^{\rm{MMSE}} = \left(\sum\limits_{i=1}^2\widehat{\vech}_i\widehat{\vech}_i^{\Htran}  + {\bf Z}\right)^{-1}\widehat{\vech}_1
\end{equation}
where ${\bf Z} =\sum\nolimits_{i=1}^2 {\bf \Phi}_i + \frac{\sigma_{\mathrm{ul}}^{2}}{\rho^{\mathrm{ul}}}{\bf{I}}_M$. 
\subsection{Downlink data transmission}\label{sec:DL_2UES}
Assume that, to transmit to \gls{ue} $i$ with $i=1,2$, the \gls{bs} uses the precoding vector ${\bf w}_i \in \mathbb{C}^{M}$, which determines the spatial directivity of the transmission and satisfies the normalization $ \Ex{}{\|  {\vecw_i} \|^2}  =1$. During \gls{dl} data transmission, the received signal $  y_1^{\mathrm{dl}} [k]\in \mathbb{C}$ at \gls{ue} $1$ over channel use $k$, where $k=1,\ldots,\ndl$, is
\begin{IEEEeqnarray}{lCl}
  y_1^{\mathrm{dl}} [k] = \herm{\vech}_1 {\bf w}_1x_1^{\mathrm{dl}}[k] + \herm{\vech}_1 {\bf w}_2x_2^{\mathrm{dl}}[k] + z_1^{\mathrm{dl}}[k]  \label{eq:simo_channel_dl}
\end{IEEEeqnarray}
where $x_i^{\mathrm{dl}}[k] \sim \jpg (0, \rho^{\mathrm{dl}})$ is the data signal intended for \gls{ue} $i$ and $z_1^{\mathrm{dl}} [k]\sim \jpg (0, \sigma_{\mathrm{dl}}^{2})$ is the receiver noise at \gls{ue} $1$. 
Again, we can put~\eqref{eq:simo_channel_dl} in the same form as~\eqref{eq:simplified_channel} by setting $v[k] = y_1^{\mathrm{dl}}[k]$, $q[k] = x_1^{\mathrm{dl}}[k]$, $g=\herm{\vech}_1 {\bf w}_1$ and $z[k] = \herm{\vech}_1 {\bf w}_2x_2^{\mathrm{dl}}[k] + z_1^{\mathrm{dl}}[k]$.
Note that, given $\{{\bf h}_1, {\bf w}_1, {\bf w}_2\}$, the random variables $\{z[k] : k=1,\ldots,\ndl\}$ are conditional \iid and $z[k]\sim \jpg(0, \sigma^2)$ with $\sigma^2 = \rho^{\mathrm{dl}}\abs{\herm{\vech}_1 {\bf w}_2}^2 + \sigma\sub{dl}^2$.

Since no pilots are transmitted in the \gls{dl}, the \gls{ue} does not know the precoded channel $g = \herm{\vech}_1 {\bf w}_1$ in \eqref{eq:simo_channel_dl}. 
Instead, we assume that the \gls{ue} has access its expected value $\Ex{}{\herm{\vech}_1 {\bf w}_1}$ and uses this quantity to perform mismatched \gls{snn} decoding. 
Specifically, we have that $\widehat{g} = \Ex{}{\herm{\vech}_1{\bf w}_1}$ and 
\begin{equation}\label{eq:mismatched_snn_decoder-donwlink}
  \widehat{\bf x}_1^{\mathrm{dl}}=\argmin_{\widetilde{\bf x}_1^{\mathrm{dl}} \in \setC^\mathrm{dl}} \vecnorm{{\bf y}_1^{\mathrm{dl}}-\widehat{g}\widetilde{\bf x}_1^{\mathrm{dl}}}^2
\end{equation}
with ${\bf y}_1^{\mathrm{dl}} = [y_1^{\mathrm{dl}}[1],\ldots,y_1^{\mathrm{dl}}[\ndl]]^{\Ttran}$ and $\widetilde{\bf x}_1^{\mathrm{dl}} = [\widetilde{x}_1^{\mathrm{dl}}[1],\ldots, \widetilde{x}_1^{\mathrm{dl}}[\ndl]]^{\Ttran}$. Obviously, channel hardening is critical for this choice to result in good performance~\cite[Sec. 2.5.1]{bjornson19}. Since $\widehat{g}=\Ex{}{\herm{\vech}_1 {\bf w}_1}$ is deterministic, the error probability at \gls{ue} $1$ in the \gls{dl} can be evaluated as follows:
\begin{IEEEeqnarray}{lCl}
  \epsilon_1^{\mathrm{dl}} \leq \Ex{}{\prob{\sum_{k=1}^{\ndl} \imath_s(y_1^{\mathrm{dl}}[k],x_1^{\mathrm{dl}}[k]) \leq \log\frac{m-1}{u} \Bigg\given g, \sigma^2 }}.\nonumber\\\label{eq:rcus_tail_simo_dl}
\end{IEEEeqnarray}

Similarly to~\eqref{eq:rcus_tail_simo_ul}, the saddlepoint approximation in Theorem~\ref{thm:saddlepoint} can be used to evaluate the conditional probability in~\eqref{eq:rcus_tail_simo_dl} efficiently. 

Similar to the UL, the upper bound \eqref{eq:rcus_tail_simo_dl} holds for any precoder vector that is selected on the basis of the channel estimates available at the \gls{bs}. 
Different precoders yield different tradeoffs between the error probability achievable at the \glspl{ue}.
A common heuristic comes from UL-DL duality~\cite[Sec.~4.3.2]{bjornson19}, which suggests to choose the precoding vectors ${\bf w}_i$ as the following function of the combining vectors: ${\bf w}_i= {{\bf u}_i}/{\sqrt{\Ex{}{\vecnorm{{\bf u}_i}^2}}}$.
By selecting ${\bf u}_i$ as one of the uplink combining schemes described earlier, the
corresponding precoding scheme is obtained; that is, ${\bf u}_i = {\bf u}_i^{\rm{MR}}$ yields \gls{mr} precoding and ${\bf u}_i = {\bf u}_i^{\rm{MMSE}}$ yields \gls{mmse} precoding. 
\subsection{Numerical Analysis}\label{sec:numerical_analysis}
In this section, we use the finite blocklength bound in Theorem~\ref{thm:rcus} to study the impact of imperfect \gls{csi}, pilot contamination, and spatial correlation in both \gls{ul} and \gls{dl}. We assume that the $K = 2$ UEs are within a square area of $75\,\text{m}\times 75 \,\text{m}$, with the \gls{bs} at the center of the square. 
The \gls{bs} is equipped with a horizontal uniform linear array (ULA) with antenna elements separated by half a wavelength.
The antennas and the \glspl{ue} are located in the same horizontal plane, thus the azimuth angle is sufficient to determine the directivity.
We assume that the scatterers are uniformly distributed in the angular interval $[\varphi_{i} -\Delta, \varphi_{i} + \Delta]$, where  $\varphi_{i}$ is the nominal angle-of-arrival (AoA) of \gls{ue} $i$ and $\Delta$ is the angular spread. 
Hence, the $(m_1,m_2)$th element of ${\bf R}_i$ is equal to \cite[Sec.~2.6]{bjornson19}
\begin{align}\label{eq:2DChannelModel}
\left[ {\bf R}_{i} \right]_{m_1,m_2} =\frac{\beta_{i}}{2\Delta} \int_{-\Delta}^{\Delta}{ e^{\mathsf{j} \pi(m_1-m_2) \sin(\varphi_{i} + {\bar \varphi}) }}d{\bar \varphi}.
\end{align}
We assume $\Delta = 25^\circ$ and let the large-scale fading coefficient, measured in \dB, be 
\begin{equation}\label{eq:pathloss-coefficient}
\beta_{i} \,  = -35.3 - 37.6\log_{10}\left( \frac{d_{i}}{1\,\text{m}} \right)
\end{equation}
where $d_{i}$\, is the distance between the \gls{bs} and \gls{ue} $i$. 
The communication takes place over a $20$\,MHz bandwidth with a total receiver noise power of $\sigma_{\mathrm{ul}}^2 = \sigma_{\mathrm{dl}}^2=-94$\,dBm (consisting of thermal noise and a noise figure of $7$\,dB in the receiver hardware) at both the \gls{bs} and \glspl{ue}.
The \gls{ul} and \gls{dl} transmit powers are equal and given by $\rho^{\mathrm{ul}}= \rho^{\mathrm{dl}} = 10\,\text{mW}$. 
We assume a total of $n = 300$ channel uses, out of which $\np$ channel uses are allocated for pilot transmission and $\nul =\ndl=(n-\np)/2$ channel uses are assigned to the \gls{ul} and \gls{dl} data transmissions, respectively.
In each data-transmission phase, $b=160$ information bits are to be conveyed.
These parameters are in agreement with the stringent low-latency setups described in \cite[App. A.2.3.1]{3GPP22.104}.
\begin{figure}[t]    
  \centering
  \begin{subfigure}{0.45\textwidth}
    \centering
    \includegraphics[width=\textwidth]{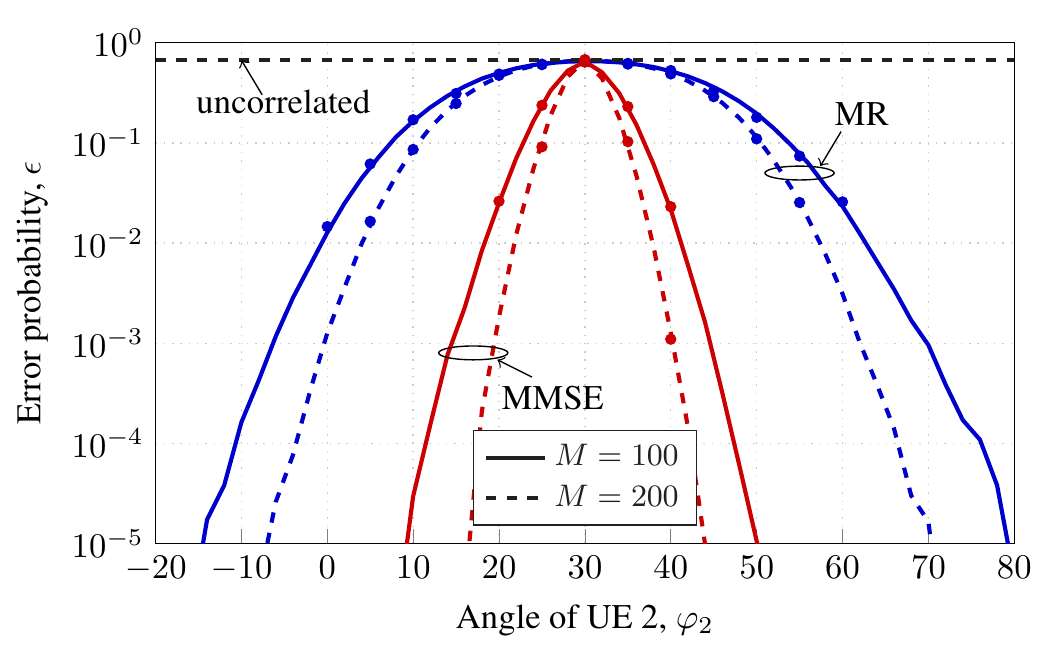}
    \caption{Uplink transmission.}
    \label{fig:angles_UL} 
  \end{subfigure}
  \hspace{3mm}
  \begin{subfigure}{0.45\textwidth}
    \centering
    \includegraphics[width=\textwidth]{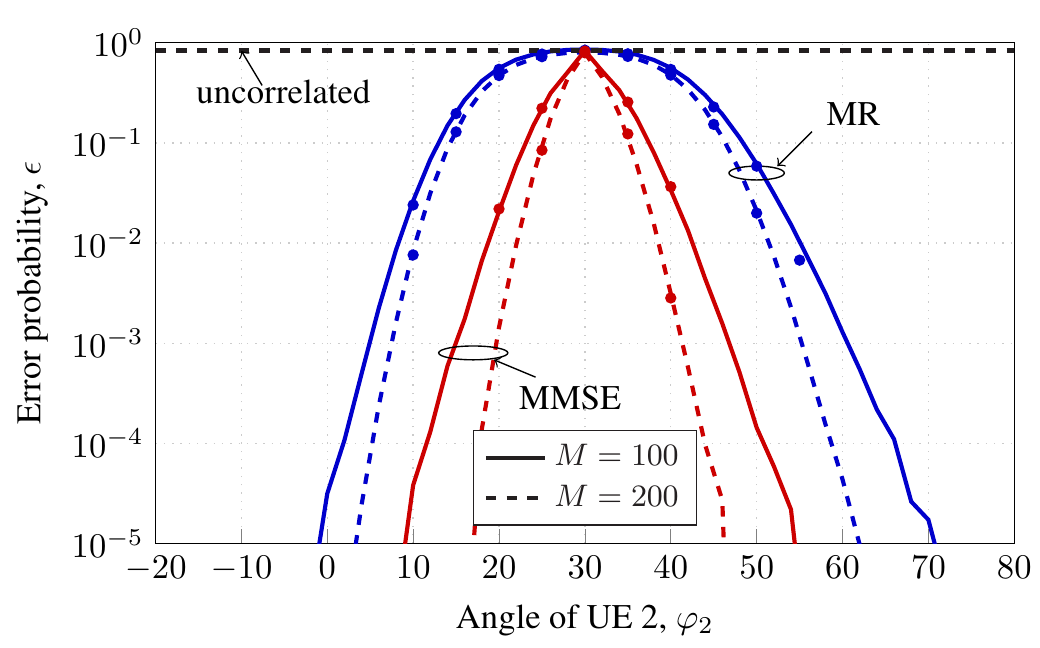}
    \caption{Downlink transmission.}
    \label{fig:angles_DL} 
  \end{subfigure}
  \caption{Average error probability $\epsilon$ for \gls{ue} $1$ versus the nominal angle of \gls{ue} $2$ when $\vecphi_1=\vecphi_2$. 
          Here, $\rho^{\text{ul}}=\rho^{\text{dl}}=10$ \dB{m}, $\Delta=25^\circ$, $\varphi_1=30^\circ$, $b=160$, $\np=2$, and $n=300$. The curves are obtained using the saddlepoint approximation; the circles indicate the values of the RCUs bound, computed directly via~\eqref{eq:rcus_fading}.}
        \label{fig:secIII_angles}
  \end{figure}

Fig.~\ref{fig:secIII_angles} shows the \gls{ul} and \gls{dl} error probability $\epsilon$ of \gls{ue} $1$ with \gls{mr} and \gls{mmse} combining, when the two \glspl{ue} use the same pilot sequence (i.e., pilot contamination is present) and $M = 100$ or $200$. The uncorrelated Rayleigh-fading case where ${\bf R}_i=\beta_i{\bf I}_M$, $i=1,2$, is also reported as reference.
The nominal angle of \gls{ue} $1$ is fixed at $\varphi_1= 30^\circ$ while the angle of \gls{ue} $2$ varies from $-20^\circ$ to $80^\circ$. 
We let $d_1=d_2=36.4$~m, which leads to $\beta_1=\beta_2=-94 \dB$. Fig.~\ref{fig:secIII_angles} reveals that a low error probability can be achieved if the \glspl{ue} are well-separated in the angle domain, even when the channel estimates are affected by pilot contamination. 
\gls{mmse} combining/precoding achieves a much lower error probability for a given angle separation.
These results are in agreement with the findings reported in the asymptotic regime of large packet size in~\cite{Sanguinetti20,Sanguinetti18}.

Fig.~\ref{fig:secIII_angles} shows that the error probability with \gls{mr} combining in the \gls{ul} is worse than that of \gls{mr} precoding in the \gls{dl}. 
This phenomenon can be clarified by comparing the input-otput relations in~\eqref{eq:simo_channel_ul} and~\eqref{eq:simo_channel_dl} for the case of perfect \gls{csi} at both \gls{bs} and \glspl{ue}.
Specifically, when the desired signal experiences a deep fade, the magnitude of the \gls{ul} interference is unaffected whereas the \gls{dl} interference becomes small. 
This results in a larger error probability in the \gls{ul} compared to the \gls{dl}.
The same argument holds also for the case of imperfect \gls{csi} with and without pilot contamination. Note that this phenomenon does not occur when \gls{mmse} combining/precoding is used.
On the contrary, with \gls{mmse} combining/precoding the \gls{dl} performs slightly worse than the \gls{ul} because \gls{dl} decoding relies on channel hardening. 

\begin{figure}[t]     
  \centering
  \begin{subfigure}{0.45\textwidth}
    \centering
    \includegraphics[width=\textwidth]{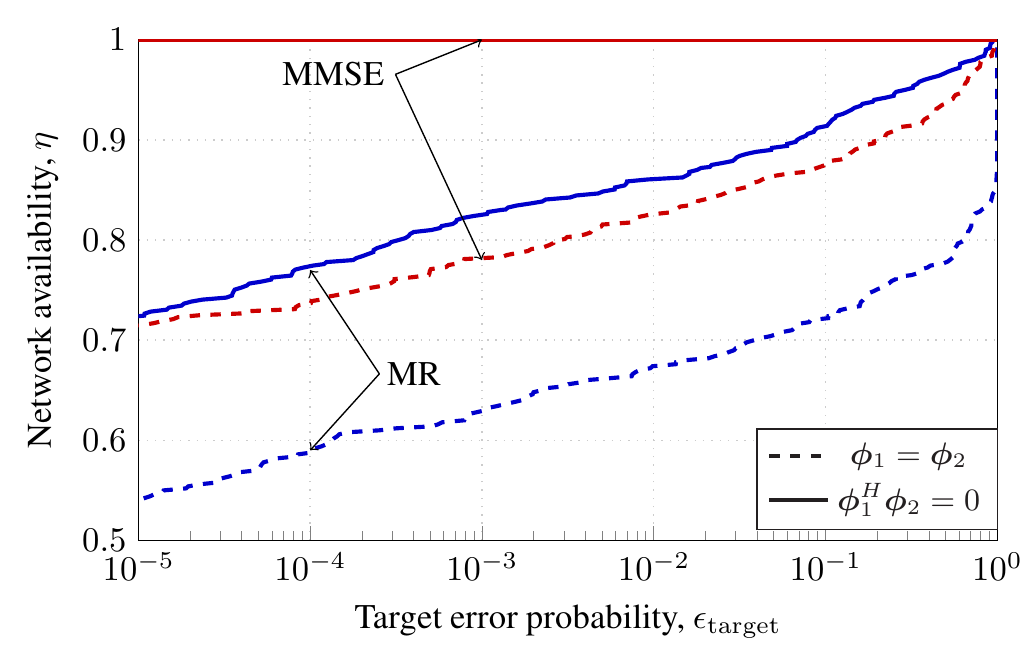}
    \caption{Uplink transmission.}
    \label{fig:secIII_CDFa} 
  \end{subfigure}
  \hspace{3mm}
  \begin{subfigure}{0.45\textwidth}
    \centering
    \includegraphics[width=\textwidth]{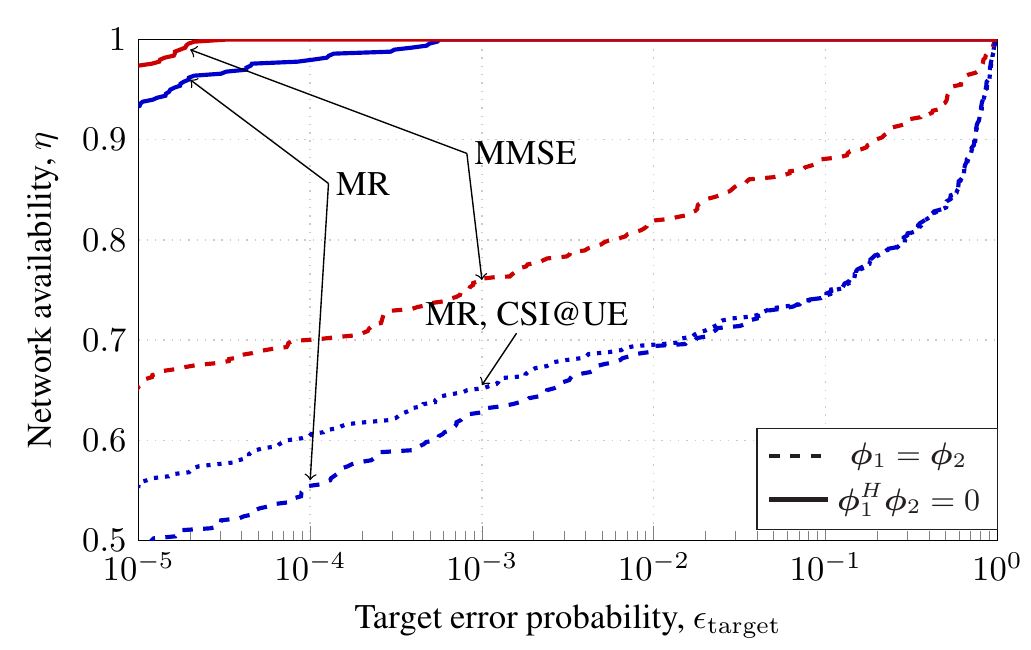}
    \caption{Downlink transmission.}
    \label{fig:secIII_CDFb} 
  \end{subfigure}
  \caption{Network availability $\eta$ with and without pilot contamination with $M=100$, $\rho^{\mathrm{ul}}=\rho^{\mathrm{dl}}=10$ \dBm, $\np=2$, $b=160$, $n=300$, and $\Delta=25^\circ$.}
       \label{fig:secIII_CDF}
\end{figure}

Assume now that the $2$ UEs are positioned independently and uniformly at random within the square area of $75\,\text{m}\times 75 \,\text{m}$, with a minimum distance from the BS of $5\,\text{m}$. 
Fig.~\ref{fig:secIII_CDF} shows the \gls{ul} and \gls{dl} network availability $\eta$ with both \gls{mr} and \gls{mmse} when $M=100$.
We define $\eta$ as
\begin{IEEEeqnarray}{rCl}\label{eq:network_avail}
 \eta = \prob{\epsilon \leq \epsilon\sub{target}}
\end{IEEEeqnarray}
and represents the probability that the target error probability $\epsilon\sub{target}$ is achieved on a link between a randomly positioned \gls{ue} and its corresponding \gls{bs}, in the presence of randomly positioned interfering \glspl{ue} (in this case, just one).
Note that the error probability $\epsilon$ is averaged with respect to the small-scale fading and the additive noise, given the \glspl{ue} location, whereas the network availability is computed with respect to the random \glspl{ue} locations. 
We consider both the scenario in which the \glspl{ue} use orthogonal pilot sequences, i.e., $\vecphi_1^{\Htran} \vecphi_2 = 0$, and the one in which $\vecphi_1 = \vecphi_2$.

The results of Fig.~\ref{fig:secIII_CDF} show that pilot contamination reduces significantly the network availability irrespective of the processing scheme. \gls{mr} performs better in the DL than in the UL when orthogonal pilot sequences are used.
This is in agreement with what stated when discussing Fig.~\ref{fig:secIII_angles}.
However, in the case of pilot contamination, the \gls{ul} achieves better performance than the \gls{dl} when the \gls{ue} relies on channel hardening (and slightly worse performance than the \gls{dl} when the \gls{ue} has access to perfect \gls{csi}).
Note that this does not contradict what stated after Fig.~\ref{fig:secIII_angles}.
Indeed, due to the random \gls{ue} placements, the correlation matrix may have low rank.
This affects channel hardening and, consequently, results in a deterioration of the \gls{dl} performance.
For \gls{mmse} processing, the \gls{ul} is always superior to the \gls{dl} because the \gls{dl} relies on channel hardening.
\subsection{Asymptotic Analysis as $M\to\infty$}\label{sec:asymptotics}
It is well known that, for spatially uncorrelated Rayleigh fading channels, the interference caused by pilot contamination limits the spectral efficiency of Massive MIMO in the large-blocklength ergodic setup as $M\to \infty$ and the number of UEs $K$ is fixed, for both \gls{mr} and \gls{mmse} combining/precoding~\cite{Marzetta10,Hoydis13}. 
However, it was recently shown in~\cite{Sanguinetti18} that Massive MIMO with \gls{mmse} combining/precoding is not asymptotically limited by pilot contamination when the spatial correlation exhibited by practically relevant channels is taken into consideration. 

We show next that a similar conclusion holds for the \emph{average error probability} in the \emph{finite-blockength} regime when \mbox{$M \to \infty$} and $K=2$.\footnote{We consider the case $K=2$ for simplicity, although a similar result can be obtained for arbitrary $K$ using the same approach.} 
Specifically, we prove that, in the presence of spatial correlation, the error probability vanishes as $M\to\infty$, provided that \gls{mmse} combining/precoding is used.
To this end, we will proceed similarly as in~\cite{Sanguinetti18} and make the following two assumptions.
\begin{assumption}\label{assumption_1}
For $i=1,2$, $\liminf_M \frac{1}{M}\trace\lro{{\bf R}_i} > 0$ and $\limsup_{M} \vecnorm{\randvecr_i}_2 < \infty$. 
\end{assumption}
\begin{assumption}\label{assumption_2}
For $(\lambda_1, \lambda_2)\in\reals^2$  and $i=1,2$,
\begin{IEEEeqnarray}{lCl} 
  \liminf_M \inf_{\lrbo{(\lambda_1, \lambda_2):\lambda_i=1}} \frac{1}{M} \vecnorm{\lambda_1 {\bf R}_1 + \lambda_2 {\bf R}_2}_F^2 > 0. \label{eq:ass2}\IEEEeqnarraynumspace
\end{IEEEeqnarray}
\end{assumption}
The first condition in Assumption~\ref{assumption_1} implies that the array gathers an amount of signal
energy that is proportional to $M$.
The second condition implies that
the increased signal energy is spread over many spatial dimensions, i.e., the rank of ${\bf R}_i$ must be proportional to $M$.
These two conditions are commonly invoked in the asymptotic analysis of Massive MIMO~\cite{Hoydis13}. 
Assumption~\ref{assumption_2} requires ${\bf R}_1$ and ${\bf R}_2$ to be asymptotically linearly independent~\cite{Sanguinetti20}. 

In Theorem~\ref{thm:MRC_asymptotic} below, we establish that, with \gls{mr} combining, the probability of error vanishes as $M\to \infty$ if the two \glspl{ue} transmit orthogonal pilot sequences.
However, it converges to a positive constant if they share the same pilot sequence.
\begin{theorem}\label{thm:MRC_asymptotic}
  Let $c>0$ be a positive real-valued scalar.
  If MR combining is used with ${\bf u}_1^{\rm{MR}} = \frac{1}{M}\widehat{\vech}_1$, then under Assumption~\ref{assumption_1}, 
  \begin{IEEEeqnarray}{lCl}
    \lim_{M\to\infty}\epsilon_1^{\mathrm{ul}} & = & 0 , \text{ if } \vecphi_1^{\Htran} \vecphi_2 = 0, \\
    \lim_{M\to\infty}\epsilon_1^{\mathrm{ul}} & = & c , \text{ if } \vecphi_1 = \vecphi_2.
  \end{IEEEeqnarray}
\end{theorem}
\begin{IEEEproof}
See Appendix~C.
\end{IEEEproof}

Next, we show that, if \gls{mmse} combining is used, the error probability vanishes as $M\to \infty$ even in the presence of pilot contamination.
\begin{theorem}\label{thm:MMSE_asymptotic}
  If \gls{mmse} combining is used with ${\bf u}_1^{\rm{MMSE}}$ given by~\eqref{eq:MMSE_combiner}, then under Assumption~\ref{assumption_1} and Assumption~\ref{assumption_2}, the average error probability $\epsilon_1^{\mathrm{ul}}$ goes to zero as $M\rightarrow \infty$, both when $\vecphi_1^{\Htran} \vecphi_2 = 0$ and when  $\vecphi_1 = \vecphi_2$.
\end{theorem}
\begin{IEEEproof}
The proof is given in Appendix~D.
It makes use of the asymptotic analysis presented in \cite[App. B]{Sanguinetti18} to show that $y_1^{\text{ul}}
  \asymp x_1^{\text{ul}}$ as $M\to \infty$, even in the presence of pilot contamination. 
  Once this is proved, the result follows by applying Lemma~\ref{lem:inf-snr} from Section~\ref{sec:fbl-intro}.
\end{IEEEproof}

 \begin{figure}[t]     
  \centering
  \begin{subfigure}{0.45\textwidth}
    \centering
    \includegraphics[width=\textwidth]{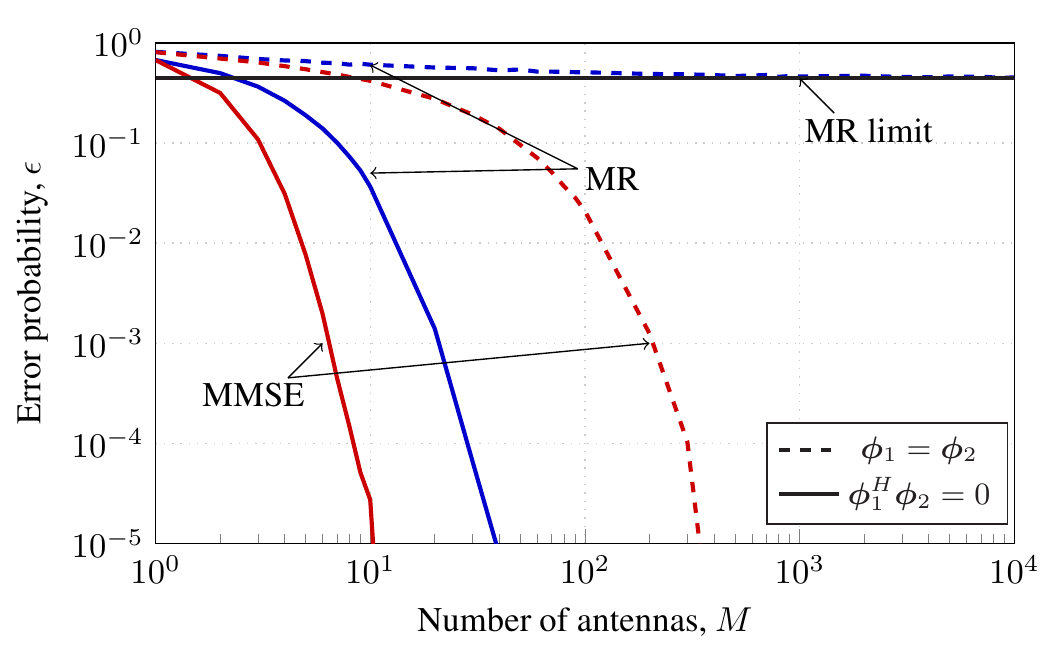}
    \caption{Uplink transmission.}
    \label{fig:asymp_UL} 
  \end{subfigure}
  \hspace{3mm}
  \begin{subfigure}{0.45\textwidth}
    \centering
    \includegraphics[width=\textwidth]{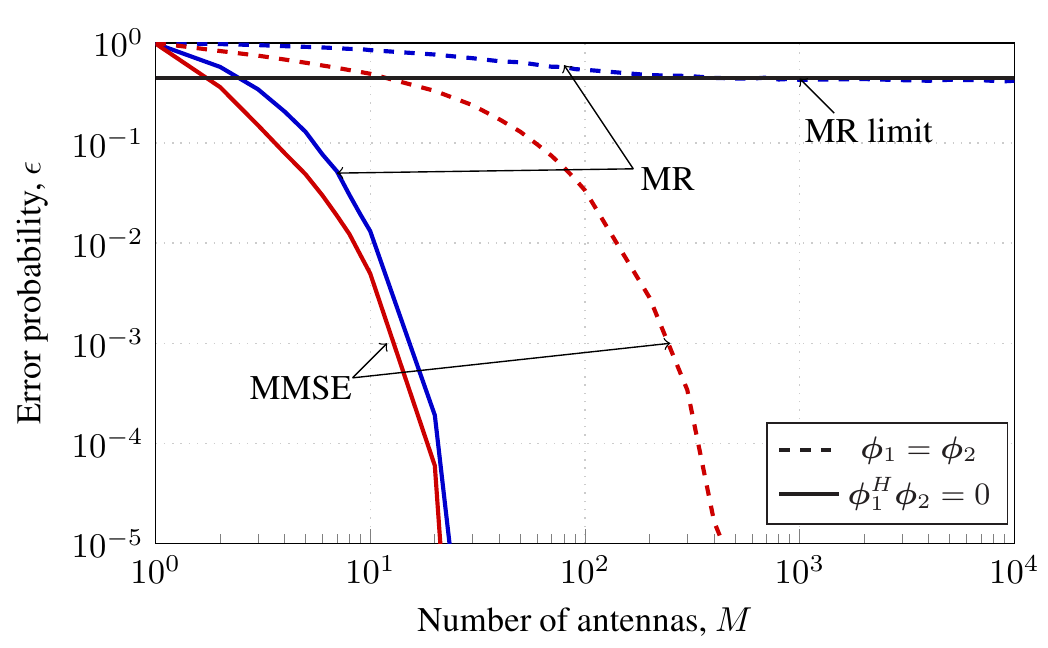}
    \caption{Downlink transmission.}
    \label{fig:asymp_DL} 
  \end{subfigure}
  \caption{Average error probability $\epsilon$ of \gls{ue} $1$ versus number of antennas $M$ with and without pilot contamination.
  Here, $\rho^{\mathrm{ul}}=\rho^{\mathrm{dl}}=10$ \dBm, $\np=2$, $b=160$, $n=300$, $\Delta=25^\circ$, $\varphi_1=30^\circ$, and $\varphi_2=40^\circ$.}
       \label{fig:secIII_asymp}
  \end{figure}

Note that Theorem~\ref{thm:MRC_asymptotic} and Theorem~\ref{thm:MMSE_asymptotic} can be extended to the \gls{dl} with a similar methodology. 
Details are omitted due to space limitations.

To validate the asymptotic analysis provided by Theorems~\ref{thm:MRC_asymptotic} and~\ref{thm:MMSE_asymptotic} and to quantify the impact of pilot contamination for values of $M$ of practical interest, we numerically evaluate the UL error probability when the $2$ \glspl{ue} transmit at the same power, are at the same distance from the \gls{bs}, and use the same pilot sequence. 
Furthermore, we assume that their nominal angles are $\varphi_1 = 30^\circ$ and $\varphi_2 = 40^\circ$. 
Note that the angle between the two UEs is small. 
Hence, we expect pilot contamination to have a significant impact on the error probability. 
As in Fig.~\ref{fig:secIII_angles}, we assume that $\sigma\sub{ul}^2 = \sigma\sub{dl}^2 = -94 \dBm$ and that the \glspl{ue} are located $36.4$ m away from the \gls{bs} so that $\beta_1=\beta_2=-94 \dB$.
In Fig.~\ref{fig:asymp_UL}, we illustrate the average error probability as a function of $M$ with \gls{mr} and \gls{mmse}. 
We see that, in the presence of pilot contamination, the error probability with \gls{mr} converges to a nonzero constant as $M$ grows, in accordance with Theorem~\ref{thm:MRC_asymptotic}. 
In contrast, the error probability with \gls{mmse} goes to $0$ as $M\to\infty$, in accordance with Theorem~\ref{thm:MMSE_asymptotic}. 
However, a comparison with the orthogonal-pilot case reveals that, for fixed $M$, pilot contamination has a significant impact on the error probability of \gls{mmse}.
As shown in Fig.~\ref{fig:asymp_DL}, similar conclusions can be drawn for the \gls{dl}. 

\section{Massive MIMO Network}\label{sec:mimo}
We will now extend the analysis in Section~\ref{sec:simo} to a Massive MIMO network with $L$ cells, each comprising a BS with $M$ antennas and $K$ UEs. We denote by ${\vech}_{li}^{j} \sim \jpg( {\veczero}_{M}, {\bf R}_{li}^{j} )$ the channel between UE~$i$ in cell~$l$ and the BS in cell~$j$. The $\np$-length pilot sequence of \gls{ue}~$i$ in cell~$j$ is denoted by the vector $\bphiu_{ji} \in \mathbb{C}^{\np}$ and satisfies $\| \bphiu_{ji} \|^2  = \np$. We assume that the $K$ UEs in a cell use mutually orthogonal pilot sequences and these pilot sequences are reused in a fraction $1/f$ of the $L$ cells with $\np = K f$. The channel vectors are estimated using the \gls{mmse} estimator given in \cite[Sec.~3.2]{bjornson19}.
 \subsection{Uplink}\label{sec:ulDataPhase}
The data signal from \gls{ue}~$i^\prime$ in cell~$l$ over an arbitrary time instant $k$ is denoted by $x_{li^\prime}^{\rm ul}[k] \sim \jpg({0}, \rho^{\rm ul})$, with $\rho^{\rm ul}$ being the transmit power. To detect $x_{ji}^{\rm ul}[k]$, \gls{bs}~$j$ selects the combining vector ${\vecu}_{ji} \in \mathbb{C}^{M}$, which is multiplied with the received signal ${\vecr}_{j}^{\rm ul}[k]$ to obtain 
\begin{IEEEeqnarray}{lCl}
\!\!y_{ji}^{\rm ul}[k] &=& {\vecu}_{ji}^{\Htran} {\vecr}_{j}^{\rm ul}[k] =  \overbrace{\underbrace{\vphantom{\sum_{i^\prime=1,i^\prime\ne i}^{K} }{\vecu}_{ji}^{\Htran}{\vech}_{ji}^{j} x_{ji}^{\rm ul}[k]}_{\textrm{Desired signal}}}^{gq[k]} + \overbrace{\underbrace{ \sum_{i^\prime=1,i^\prime\ne i}^{K} {\vecu}_{ji}^{\Htran}{\vech}_{ji^\prime}^{j} x_{ji^\prime}^{\rm ul}[k]}_{\textrm{Intra-cell interference}}}^{z[k]} \nonumber\\ 
 &&{}+ \overbrace{\underbrace{\sum_{l=1,l \neq j}^{L} \sum_{i^\prime=1}^{K} {\vecu}_{ji}^{\Htran}{\vech}_{li^\prime}^{j} x_{li^\prime}^{\rm ul}[k]}_{\textrm{Inter-cell interference}}}^{z[k]} + \overbrace{\underbrace{\vphantom{\sum_{i^\prime=1,i^\prime\ne i}^{K} } {\vecu}_{ji}^{\Htran}{\vecz}_{j}^{\rm ul}[k]}_{\textrm{Noise}}}^{z[k]}\label{eq:vector-channel-UL-processed}
\end{IEEEeqnarray}
for $k=1,\ldots,\nul$. We note that~\eqref{eq:vector-channel-UL-processed} can be put in the same form as~\eqref{eq:simplified_channel} if we set $v[k] = y_{ji}^{\rm ul}[k]$, $q[k] = x_{ji}^{\rm ul}[k]$, $g = {\vecu}_{ji}^{\Htran}{\vech}_{ji}^{j}$, $\widehat{g}={\vecu}_{ji}^{\Htran}{\widehat{\vech}}_{ji}^{j}$, and $z[k] = \sum_{i^\prime=1,i^\prime\ne i}^{K} {\vecu}_{ji}^{\Htran}{\vech}_{ji^\prime}^{j} x_{ji^\prime}^{\rm ul}[k] + \sum_{l=1,l \neq j}^{L} \sum_{i^\prime=1}^{K} {\vecu}_{ji}^{\Htran}{\vech}_{li^\prime}^{j} x_{li^\prime}^{\rm ul}[k]+{\vecu}_{ji}^{\Htran}{\vecz}_{j}^{\rm ul}[k]$.
Given all channels and combining vectors, the random variables $\{z[k] : k=1,\ldots,\nul\}$ are conditionally \iid and $z[k]\sim \jpg(0, \sigma^2)$ with $\sigma^2 = \sigma\sub{ul}^2\vecnorm{\vecu_{ji}}^2 + \rho^{\mathrm{ul}}\sum_{i^\prime=1, i^\prime\neq i}^K \abs{\herm{\vecu}_{ji} \vech_{ji^{\prime}}^j}^2 + \rho^{\mathrm{ul}}\sum_{l=1, l\neq j}^L \sum_{i^{\prime}=1}^K \abs{\herm{\vecu}_{ji} \vech_{li^{\prime}}^j}^2$.
An upper bound on the error probability $\epsilon_{ji}^{\rm ul}$ then follows by applying~\eqref{eq:rcus_tail} in Theorem~\ref{thm:rcus} and then by averaging over $g$, $\widehat{g}$ and $\sigma^2$.  
This bound holds for any choice of ${\bf v}_{ji}$. 
In the numerical results, we will consider multicell MMSE and MR combining.
\subsection{Downlink}\label{sec:dlDataPhase}
The \gls{bs} in cell~$j$ transmits the \gls{dl} signal ${\vecx}_j^{\rm dl}[k] = \sum_{ji^\prime=1}^{K} {\vecw}_{ji^\prime} x_{ji^\prime}^{\rm dl}[k]$
where  $x_{ji^\prime}^{\rm dl}[k] \sim \jpg(0,\rho^{\rm dl})$ is the \gls{dl} data signal intended for \gls{ue}~$i^\prime$ in cell $j$ over the time index $k$, assigned to a precoding vector $ {\vecw}_{ji^\prime} \in \mathbb{C}^{M}$ that satisfies $ \|  {\vecw}_{ji^\prime} \|^2  =1$ so that $\rho^{\rm dl}$ represents the transmit power.
The received signal $y_{ji}^{\rm dl}[k] \in \mathbb{C}$ for $k=1,\ldots, \ndl$ at \gls{ue}~$i$ in cell~$j$ is given by
\begin{IEEEeqnarray}{lCl}
y_{ji}^{\rm dl}[k] &=& \overbrace{\underbrace{   \vphantom{\sum_{i^\prime=1,i^\prime\ne i}^{K} }   ( {\vech}_{ji}^{j})^{\Htran} {\vecw}_{ji} x_{ji}^{\rm dl}[k]}_{\textrm{Desired signal}}}^{gq[k]} + \overbrace{\underbrace{\sum_{i^\prime=1,i^\prime\ne i}^{K}  ({\vech}_{ji}^{j})^{\Htran} {\vecw}_{ji^\prime} x_{ji^\prime}^{\rm dl}[k]}_{\textrm{Intra-cell interference}}}^{z[k]} \nonumber\\
&&{} + \overbrace{\underbrace{\sum_{l=1, l\neq j}^{L} \sum_{i^\prime=1}^{K}  ({\vech}_{ji}^{l})^{\Htran} {\vecw}_{li^\prime} x_{li^\prime}^{\rm dl}[k]}_{\textrm{Inter-cell interference}}}^{z[k]} + \overbrace{\underbrace{ \vphantom{\sum_{i^\prime=1,i^\prime\ne i}^{K} }z_{ji}^{\rm dl}[k]}_{\textrm{Noise}}}^{z[k]}\label{eq:downlink-signal-model}
\end{IEEEeqnarray}
where $z_{ji}^{\rm dl} [k]\sim \jpg(0,\sigma\sub{dl}^2)$ is the receiver noise.
The desired signal to UE $i$ in cell $j$ propagates over the precoded channel $g_{ji} = ( {\vech}_{ji}^{j})^{\Htran} {\vecw}_{ji}$.
The \gls{ue} does not know $g_{ji}$ and relies on channel hardening to approximate it with its mean value $\Ex{}{g_{ji}} =\Ex{}{( {\vech}_{ji}^{j})^{\Htran} {\vecw}_{ji}}$.
As in the UL, we note that~\eqref{eq:downlink-signal-model} can be put in the same form as~\eqref{eq:simplified_channel} if we set $v[k] = y_{ji}^{\rm dl}[k]$, $q [k]= x_{ji}^{\rm dl}[k]$, $g = ( {\vech}_{ji}^{j})^{\Htran} {\vecw}_{ji}$, $\widehat{g}=\Ex{}{( {\vech}_{ji}^{j})^{\Htran} {\vecw}_{ji}}$, and $z[k] = \sum_{i^\prime=1,i\ne i}^{K}  ({\vech}_{ji}^{j})^{\Htran} {\vecw}_{ji^\prime} x_{ji^\prime}^{\rm dl}[k]+ \sum_{l=1, l\neq j}^{L} \sum_{i^\prime=1}^{K}  ({\vech}_{ji}^{l})^{\Htran} {\vecw}_{li^\prime} x_{li^\prime}^{\rm dl}[k] + z_{ji}^{\rm dl}[k]$.
Given all channels and precoding vectors, the random variables $\{z[k] : k=1,\ldots,\ndl\}$ are conditionally \iid and $z[k]\sim \jpg(0, \sigma^2)$ with $\sigma^2 = \sigma\sub{dl}^2 + \rho^{\mathrm{dl}}\sum_{i^{\prime}=1, i\neq i^{\prime}}^K \abs{\herm{(\vech_{ji}^j)}\vecw_{ji^{\prime}}}^2 + \rho^{\mathrm{dl}} \sum_{l=1, l\neq j}^L \sum_{i^{\prime}=1}^K \abs{\herm{(\vech_{ji}^l)}\vecw_{li^{\prime}}}^2$.
An upper bound on the error probability $\epsilon_{ji}^{\rm dl}$ then follows by applying~\eqref{eq:rcus_tail} in Theorem~\ref{thm:rcus} and then by averaging over $g$ and $\sigma^2$.  
As for the UL, the above results hold for any choice of ${\bf w}_{ji}$. 
In the numerical simulations, we will consider both multicell MMSE and MR precoding.
\subsection{Numerical Analysis}\label{sec:numericalResult}
The simulation setup consists of $L=4$ square cells, each of size $75$ m $\times$ $75$ m, containing $K=10$ \glspl{ue} each, independently and uniformly distributed within the cell, at a distance of at least $5$ m  from the \gls{bs}.
 As in Section~\ref{sec:numerical_analysis}, we consider a horizontal ULA with $M=100$ antennas and half-wavelength spacing.
The correlation matrix and large-scale fading coefficient associated with each \gls{ue} follow the models given in~\eqref{eq:2DChannelModel} and~\eqref{eq:pathloss-coefficient}, respectively.
 Furthermore, we employ a wrap-around topology as in~\cite[Sec. 4.1.3]{bjornson19}. As in Section~\ref{sec:numerical_analysis}, we assume $n = 300$, $\nul =\ndl=(n-\np)/2$, $b=160$ and $\rho^{\mathrm{ul}} = \rho^{\mathrm{dl}}=10 \dBm$.

\begin{figure}[t]     
  \centering
  \begin{subfigure}{0.45\textwidth}
    \centering
    \includegraphics[width=\textwidth]{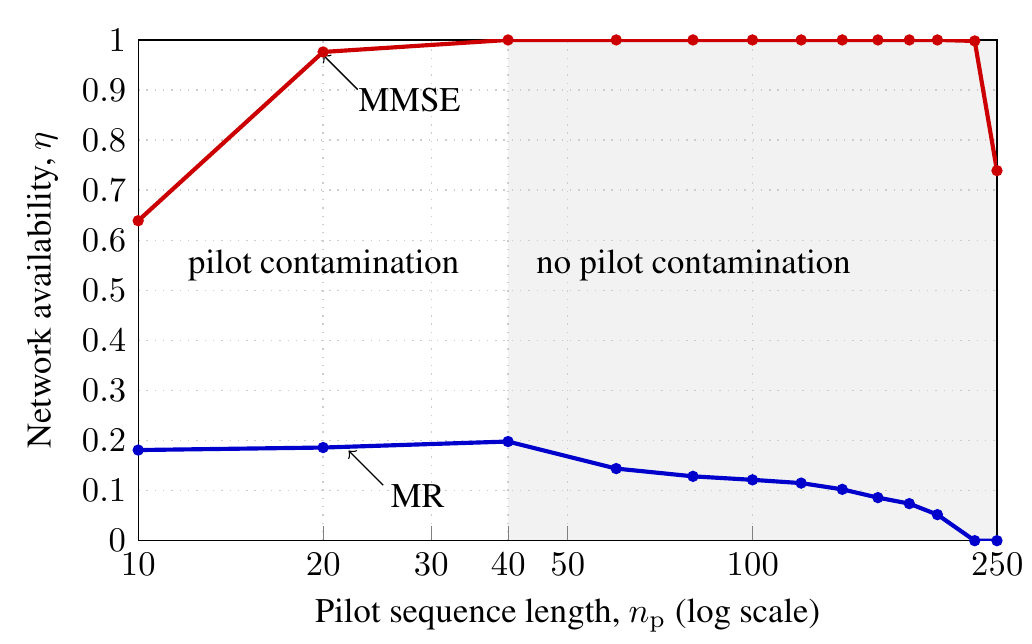}
    \caption{Uplink transmission.}
    \label{fig:pilots_UL} 
  \end{subfigure}
  \hspace{3mm}
  \begin{subfigure}{0.45\textwidth}
    \centering
    \includegraphics[width=\textwidth]{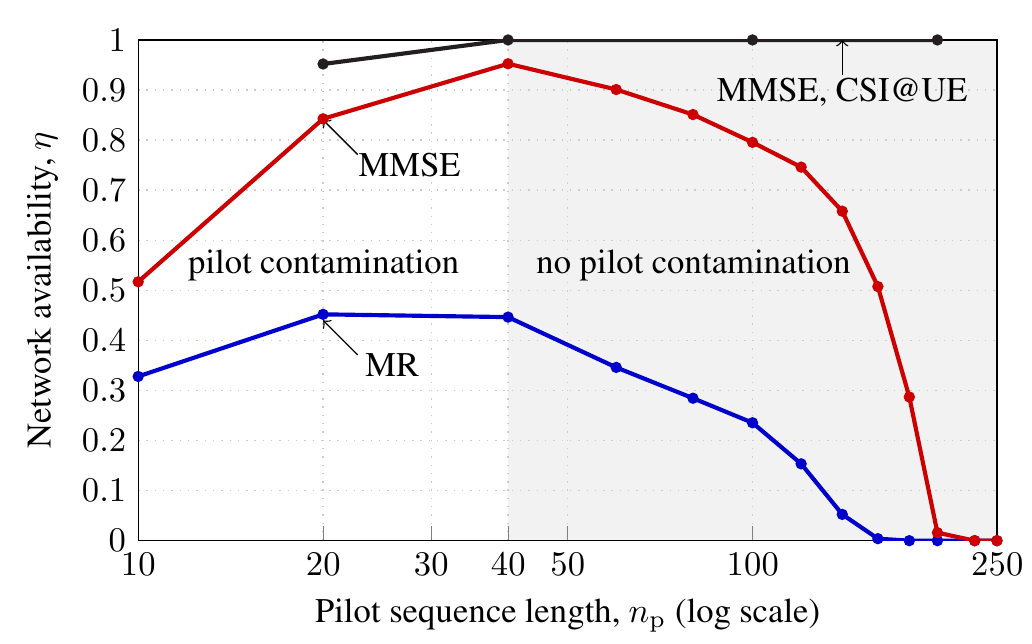}
    \caption{Downlink transmission.}
    \label{fig:pilots_DL} 
  \end{subfigure}
  \caption{Network availability for $\epsilon\sub{target} = 10^{-5}$. 
  Here, $L=4$, $K=10$, $\Delta=25^\circ$, the cell size is $75\times 75$ m, $\rho^{\mathrm{ul}} = \rho^{\mathrm{dl}}=10 \dBm$, $M=100$, $b = 160$, and $n=300$.}
       \label{fig:pilots_vs_avail}
\end{figure} 

In Fig.~\ref{fig:pilots_vs_avail}, we plot the network availability~\eqref{eq:network_avail} for a fixed $\epsilon\sub{target}=10^{-5}$ (which is in agreement with the \gls{urllc} requirements) versus the number of pilot symbols $\np = fK$, where we recall that $f$ is the pilot reuse factor. 
The results presented in Fig.~\ref{fig:secIII_CDF} suggest that pilot contamination should be avoided and that \gls{mmse} should be preferred to \gls{mr}.
The results presented in Fig.~\ref{fig:pilots_vs_avail} confirm these design guidelines.
With multicell \gls{mmse}, a network availability above $90\%$ can be achieved in UL and DL by setting a pilot reuse factor $f  = 4$ such that $\np=fK=40$. This is the minimum value of $\np$ that results in no pilot contamination in a network with $L=4$ cells.
Increasing $\np$ further has a deleterious effect on the network availability, especially in the DL. 
Indeed, the corresponding reduction in the number of channel uses $\ndl=(300-\np)/2$ available for data transmission in the DL overcomes the benefits of a more accurate \gls{csi}.
As already discussed, the difference in performance between \gls{ul} and \gls{dl} with multicell \gls{mmse} processing is due to the assumption that the \gls{ue} has no \gls{csi} and performs mismatched decoding by relying on channel hardening.
Indeed, when the UEs are provided with perfect \gls{csi} (black curve), the network availability achievable in \gls{ul} and \gls{dl} is the same.  
If needed, additional network-availability gains can be achieved by, e.g., increasing the number of BS antennas, by reducing the number of UEs that are served simultaneously, or by using scheduling to avoid serving at the same time UEs that are difficult to separate spatially via linear precoding. 
For example, in the scenario considered in Fig.~\ref{fig:pilots_DL}, a network availability above $98\%$ can be achieved by halving the number of scheduled UEs.
Finally, note that the network availability achievable with \gls{mr} is below $50\%$ even when pilot contamination is avoided. 
This implies that in practical scenarios, \gls{mr} is too sensitive to interference to achieve the low error probability targets required in \gls{urllc}. 
 
\section{Conclusions}\label{sec:conclusions} 
We presented guidelines on the design of Massive MIMO systems supporting the transmission of short information packets under the high reliability targets demanded in \gls{urllc}.
Specifically, we showed that, for a BS equipped with up to $100$ antennas, it is imperative to avoid pilot contamination and to use \gls{mmse} spatial processing in place of the computationally less intensive \gls{mr} spatial processing.
Our guidelines were based on a firm nonasymptotic bound on the error probability, which is based on recent results in finite-blocklength information theory, and applies to a realistic Massive MIMO network, with imperfect channel state information, pilot contamination, spatially correlated channels, arbitrary linear spatial processing, and randomly positioned \glspl{ue}.
We provided an accurate approximation for this bound, based on the saddlepoint method, which makes its evaluation computationally efficient for the low  error probabilities targeted in \gls{urllc}.
Finally, we showed that analyses based on performance metrics such as outage probability and normal approximation, although appealing because of the simplicity of the underlying mathematical formulas, may result in a significant underestimation of the error probability, which is clearly undesirable when designing \gls{urllc} links.
Results relied on the assumption that the channel covariance matrix is perfectly known to the receiver. 
If no such knowledge is available, the \gls{bs} can perform instead least-square channel estimation followed by regularized zero forcing, at the cost of a performance loss recently quantified in~\cite{lancho21-06a}.
 
\appendices

\section*{Appendix A - Proof of Theorem~\ref{thm:rcus}}\label{app:rcus}
Let $\vecq=[q[1],\dots,q[n]]\sim\jpg({\bf 0},\snr  {{\bf I}_n})$ be the transmitted codeword and $\vecv=[v[1],\dots,v[n]]$ be the corresponding channel output obtained via the input-output relation~\eqref{eq:simplified_channel}.
Finally, let $\bar{\vecq}=[\bar{q}[1],\dots,\bar{q}[n]]$ be a vector of \gls{iid} $\jpg(0,\snr)$ random variables, independent of both $\vecq$ and $\vecv$.
Intuitively, $\bar{\vecq}$ stands for any codeword different from the transmitted one.

A simple generalization of the random coding union bound in~\cite[Th.~16]{polyanskiy10-05a} to the mismatched \gls{snn} decoder~\eqref{eq:mismatched_snn_decoder} results in the following bound
\begin{equation}\label{eq:rcu}
  \epsilon\leq \Ex{}{\min\lefto\{1,(m-1)f(\vecq,\vecv)\right\}}
\end{equation}
where $f(\vecq,\vecv)=\Pr\{\vecnorm{\vecv-\widehat{g}\bar{\vecq}}^2\leq \vecnorm{\vecv-\widehat{g}{\vecq}}^2\given \vecq,\vecv\}$.
The bound~\eqref{eq:rcu} is obtained by observing that, when the mismatched \gls{snn} decoder~\eqref{eq:mismatched_snn_decoder} is used, an error occurs if, after being scaled by $\widehat{g}$, the codeword $\bar{\vecq}$ is closer in Euclidean distance to $\vecv$ than $\vecq$, and then by using a tightened version of the union bound. We next apply the Chernoff bound to $f(\vecq,\vecv)$ and obtain that
\begin{IEEEeqnarray}{lCl}
 f(\vecq,\vecv) &\leq& \frac{\Ex{\bar{\vecq}}{\exp\lro{-s\vecnorm{\vecv-\widehat{g}\bar{\vecq}}^2}}}{\exp\lro{-s\vecnorm{\vecv-\widehat{g}\vecq}^2}}\label{eq:chernoff_fqv}
\end{IEEEeqnarray}
for $s>0$.
Substituting~\eqref{eq:chernoff_fqv} into \eqref{eq:rcu}, we conclude that
\begin{IEEEeqnarray}{lCl}
\epsilon &\leq& \mathbb{E}\Biggl[\min\Biggl\{1,\exp\Biggl(\log(m-1) \nonumber\\
&& \qquad\qquad {} + \log\frac{\Ex{\bar{\vecq}}{\exp\lro{-s\vecnorm{\vecv-\widehat{g}\bar{\vecq}}^2}}}{\exp\lro{-s\vecnorm{\vecv-\widehat{g}\vecq}^2}}\Biggr)\Biggr\}\Biggr]\\
&=& \mathbb{E}\Biggl[\min\Biggl\{1,\exp\Biggl(\log(m-1) \nonumber\\
&& \,\,\, {} - \sum_{k=1}^n\log\frac{\exp\lro{-s|v[k] - \widehat{g}q[k]|^2}}{\Ex{\bar{q}[k]}{\exp\lro{-s|v[k] - \widehat{g}\bar{q}[k]|^2}}}\Biggr)\Biggr\}\Biggr].\label{eq:RCUs_step1} \IEEEeqnarraynumspace
\end{IEEEeqnarray}
Let now the generalized information density be defined as
\begin{equation}
 \imath_s(q[k],v[k]) = \log \frac{\exp\lro{-s|v[k] - \widehat{g}q[k]|^2}}{\Ex{\bar{q}[k]}{\exp\lro{-s|v[k] - \widehat{g}\bar{q}[k]|^2}}}.\label{eq:gen_info_dens_def}
\end{equation}
Using~\eqref{eq:gen_info_dens_def}, we can rewrite~\eqref{eq:RCUs_step1} as
\begin{IEEEeqnarray}{lCl}
 \epsilon &\leq& \mathbb{E}\Biggl[\min\Biggl\{1,\exp\Biggl(\log(m-1) \nonumber\\
 && \qquad\qquad\qquad\quad {} - \sum_{k=1}^n\imath_s(q[k],v[k])\Biggr)\Biggr\}\Biggr].\label{eq:RCUs_step2} 
 \end{IEEEeqnarray}
The desired bound~\eqref{eq:rcus_tail} follows by observing that, for every positive random variable $w$, we have that $\Ex{}{\min\lrbo{1,w}} = \prob{w \geq u}$ where $u$ is uniformly distributed on $\lrho{0,1}$.

To conclude the proof, it remains to show that the generalized information density defined in~\eqref{eq:gen_info_dens_def} can be expressed as in~\eqref{eq:simple_infodens}. 
 Since $\bar{q}[k]\sim\jpg(0,\rho)$, it follows that, for a given $v[k]$ 
\begin{IEEEeqnarray}{lCl}
 |v[k] - \widehat{g}\bar{q}[k]|^2 &\stackrel{d}{=}& \frac{|\widehat{g}|^2\rho}{2}\lro{\lro{\frac{|v[k]|\sqrt{2}}{|\widehat{g}|\sqrt{\rho}} + u_1}^2 + u_2^2} \nonumber\\
 &\stackrel{d}{=}& \frac{|\widehat{g}|^2\rho}{2} \theta\label{eq:dist_exp_denom}
\end{IEEEeqnarray}
where $u_1$ and $u_2$ are independent $\normal(0,1)$ random variables and $\theta$ follows a noncentral chi-squared distribution with $2$ degrees of freedom and noncentrality parameter $\lambda={2|v[k]|^2}/(\rho|\widehat{g}|^2)$. The \gls{mgf} of $\theta$ is given by 
\begin{equation}\label{eq:mgf}
  \Ex{}{e^{\zeta \theta}} = \frac{\exp\lro{\frac{\lambda\zeta}{1-2\zeta}}}{(1-2\zeta)}, \quad \zeta<\frac{1}{2}.
\end{equation}
Using~\eqref{eq:mgf} in~\eqref{eq:gen_info_dens_def} with $\zeta = - s|\widehat{g}|^2\rho/2$, we conclude that~\eqref{eq:gen_info_dens_def} coincides with~\eqref{eq:simple_infodens}.

\section*{Appendix B - Proof of~\eqref{eq:RoC_values_A} and~\eqref{eq:RoC_values_B}}\label{app:CGF}
In this appendix, we prove that~\eqref{eq:cond_saddle} holds for every $\zeta \in [\underline{\zeta},\overline{\zeta}]$, where $\underline{\zeta}$ and $\overline{\zeta}$ are given in~\eqref{eq:RoC_values_A} and~\eqref{eq:RoC_values_B}, respectively.
Let $q\sim \jpg(0,\rho)$ and $v=gq+z$ where $z\distas\jpg(0,\sigma^2)$, so that $v \sim \jpg(0, \sigma_v^2)$ with $\sigma_v^2 = \rho  \abs{g}^2 + \sigma^2 $
Furthermore, set
$A = s\abs{v - \widehat{g} q}^2 $ and $B= \gamma \abs{v}^2$ with $\gamma = s/(1+s \rho \abs{\widehat{g}}^2)$.
We can then rewrite the information density~\eqref{eq:simple_infodens} as\footnote{We drop the indices in $q$ and $v$ because immaterial for the proof.}
\begin{IEEEeqnarray}{lCl}
  \imath_s\lro{q,v} &=&  B-A + \log\lro{1+s\rho\abs{\widehat{g}}^2}
  \label{eq:info_dens_alt}
\end{IEEEeqnarray}
It then follows that $A$ and $B$ are dependent exponentially-distributed random variables with rate parameter $1/\beta_A$ defined in~\eqref{eq:betaA}  and $1/\beta_B$ defined in~\eqref{eq:betaB}, respectively. 
This implies that the random variable $\imath_s\lro{q,v}$ involves the difference between two dependent exponentially-distributed random variables.
Let $\Delta=B-A$. 
The \gls{pdf} of $\Delta$ is~\cite[Cor. 8]{Holm-04}
\begin{IEEEeqnarray}{lCl}
  f_{\Delta}(\delta) &=& \frac{1}{\sqrt{(\beta_B-\beta_A)^2 + 4\beta_A\beta_B(1-\nu)}} \nonumber \\
  &&\> \times \exp\lro{- \frac{\abs{\delta}\sqrt{(\beta_B-\beta_A)^2 + 4\beta_A\beta_B(1-\nu)}}{2\beta_A\beta_B(1-\nu)}}\nonumber\\ 
  && \> \times \exp\lro{\frac{\delta \lro{\beta_B-\beta_A}}{2\beta_A\beta_B(1-\nu)}}
  \label{eq:pdf_diff}
\end{IEEEeqnarray}
where $\nu= \text{Cov}\lro{A, B}/\sqrt{\text{Var}\lro{A}\text{Var}\lro{B}}$ is the correlation coefficient between $A$ and $B$. 
Using~\eqref{eq:pdf_diff}, we can express the \gls{mgf} of~$-\imath_s\lro{q,v}$ as follows: 
\begin{IEEEeqnarray}{lCl}
 \!\!\! \Ex{}{e^{-\zeta \imath_s\lro{q,v}}} &=& \frac{1}{(1+s\rho\abs{\widehat{g}}^2)^\zeta}
  \int_{-\infty}^{\infty} \exp\lro{-\zeta \delta} f_{\Delta}(\delta)\mathrm{d}\delta\nonumber\\
   &=& \frac{(1+s\rho\abs{\widehat{g}}^2)^{-\zeta}}{ 1+(\beta_B-\beta_A)\zeta - \beta_A\beta_B(1-\nu)\zeta^2}\label{eq:mgf_closed} 
\end{IEEEeqnarray}
where the last step holds for all $\zeta \in  (\underline{\zeta}, \overline{\zeta})$, with $\underline{\zeta}$ and $\overline{\zeta}$ given in \eqref{eq:RoC_values_A} and \eqref{eq:RoC_values_B}, respectively.
 The desired result in~\eqref{eq:cond_saddle} follows because the right-hand side of \eqref{eq:mgf_closed} is infinitely differentiable.

To conclude the proof, we need to show that $\nu$ in~\eqref{eq:pdf_diff} is given by~\eqref{eq:corr_coeff_SISO}.
By definition, $  \text{Cov}\lro{A,B}= \Ex{}{AB} - \Ex{}{A}\Ex{}{B}$
where
\begin{equation}\label{eq:correlation_ab}
  \Ex{}{AB}=s\gamma\Ex{}{\abs{v - \widehat{g} q}^2 \abs{v}^2}.
\end{equation}
To compute this correlation, it turns out convenient to set $x=v-\widehat{g}q$ and to express $x$ as the \gls{mmse} estimate of $v$ given $x$ plus the uncorrelated estimation error $e$:
\begin{equation}\label{eq:mmse-est-app}
  x=\alpha v+e.
\end{equation}
Here, $\alpha$ is the \gls{mmse} coefficient, given by $\alpha=\Ex{}{v^*x}/\sigma_v^2$, and $e\distas\jpg(0,\sigma_e^2)$ where $\sigma_e^2=\sigma^2_x-\abs{\Ex{}{v^*x}}^2/\sigma_v^2$, with $\sigma^2_x=\Ex{}{\abs{x}^2}=\abs{g-\widehat{g}}^2\snr+\sigma^2$.
Note that since $e$ is Gaussian and uncorrelated with $v$, then $e$ and $v$ are independent.
Using~\eqref{eq:mmse-est-app}, we can rewrite the expectation on the right-hand side of~\eqref{eq:correlation_ab} as follows:
\begin{IEEEeqnarray}{rCl}\nonumber
  \Ex{}{\abs{v - \widehat{g} q}^2 \abs{v}^2} &=& \Ex{}{\abs{x}^2\abs{v}^2}\\\nonumber
  &=& \Ex{}{\abs{\alpha v+ e}^2\abs{v}^2 }\\\nonumber
  &=&  \abs{\alpha}^2 \Ex{}{\abs{v}^4} + \Ex{}{\abs{v}^2} \Ex{}{\abs{e}^2}\IEEEeqnarraynumspace \\
  &=&  2\abs{\alpha}^2 \sigma_v^4 +  \sigma_v^2 \sigma_e^2 \label{eq:correlation-derivation-1}\\
  &=&  \abs{\Ex{}{v^*x}}^2 +\sigma_v^2 \sigma_x^2. \label{eq:correlation-derivation-2}
\end{IEEEeqnarray}
Here, in~\eqref{eq:correlation-derivation-1} we used that $\Ex{}{\abs{v}^4}=2\sigma_v^4$.
Furthermore,~\eqref{eq:correlation-derivation-2} follows by the definition of~$\alpha$ and of~$\sigma^2_e$.
Note now that  $\beta_A=s\sigma_x^2$ and $\beta_B=\gamma \sigma^2_v$.
Recall also that $\Ex{}{A} = \beta_A$, $\text{Var}\lro{A}= \beta_A^2$, $\Ex{}{B} = \beta_B$, $\text{Var}\lro{B}= \beta_B^2$.
Hence, we conclude that 
\begin{IEEEeqnarray}{rCl}
  \nu =
  \frac{s\gamma(\abs{\Ex{}{v^*x}}^2 +\sigma_v^2 \sigma_x^2) -\beta_A\beta_B}{\beta_A\beta_B}=\frac{s\gamma \abs{\Ex{}{v^*x}}^2}{\beta_A \beta_B}. \IEEEeqnarraynumspace
\end{IEEEeqnarray}
To obtain~\eqref{eq:corr_coeff_SISO}, we use that $\gamma = s/(1+s \rho \abs{\widehat{g}}^2)$ and that $\Ex{}{v^*x}=\sigma_v^2-g^*\widehat{g}\snr$.
\section*{Appendix C - Proof of Theorem~\ref{thm:MRC_asymptotic}}\label{app:MRC_asymp}
Substituting ${\bf u}_1^{\rm{MR}} = \frac{1}{M}\widehat{\vech}_1$ into~\eqref{eq:simo_channel_ul}, we obtain
\begin{equation}
y_1^{\mathrm{ul}}[k] = \frac{1}{M}\herm{\widehat{\vech}}_1\vech_1 x_1^{\mathrm{ul}}[k] +  \frac{1}{M}\herm{\widehat{\vech}}_1\vech_2 x_2^{\mathrm{ul}}[k] + \frac{1}{M}\herm{\widehat{\vech}}_1 \vecz^{\mathrm{ul}}[k]. \label{eq:app_mr_start}
\end{equation}
Under Assumption~\ref{assumption_1} and using \cite[Lem.~3]{Sanguinetti18}, we have that, in the limit ${M \to \infty}$,\footnote{Under Assumption~\ref{assumption_1}, ${\bf R}_k{\bf Q}^{-1} {\bf R}_i$ has uniformly bounded spectral norm---a result that follows from~\cite[Lem.~4]{Sanguinetti18}.}
\begin{align}
 \frac{1}{M}\herm{\widehat{\vech}}_1\vech_1 \mathop{\asymp}^{(a)} \frac{1}{M}\herm{\widehat{\vech}}_1{\widehat{\vech}}_1 \asymp \frac{1}{M}\tr({\bf{\Phi}}_1) \, \text{and} \,
\frac{1}{M}\herm{\widehat{\vech}}_1 \vecz^{\mathrm{ul}}[k] \mathop{\asymp}^{(b)} 0.\label{eq:app_mr_term1}
\end{align}
Here, ${(a)}$ and $(b)$ follow because ${\widehat{\vech}}_1$ and the pair $({\widetilde{\vech}}_1 , \vecz^{\mathrm{ul}}[k])$ are independent.
Similarly, we have that
\begin{IEEEeqnarray}{lCl}
    \frac{1}{M}\herm{\widehat{\vech}}_1\vech_2 &\asymp & 0, \text{ if } \vecphi_1 \ne  \vecphi_2 \; \text{with}\;\vecphi_1^{\Htran} \vecphi_2 = 0,\label{eq:app_mr_term2} \\
    \frac{1}{M}\herm{\widehat{\vech}}_1\vech_2 &\asymp & \frac{1}{M}\herm{\widehat{\vech}}_1{\widehat{\vech}}_2 \mathop{\asymp}^{(c)}\frac{1}{M}\tr ({\bf \Upsilon}_{12}) ,  \text{ if } \vecphi_1 = \vecphi_2\label{eq:app_mr_term3}\IEEEeqnarraynumspace
  \end{IEEEeqnarray}
  where ${(c)}$ follows from the fact that ${\widehat{\vech}}_1$ and ${\widehat{\vech}}_2$ are correlated under pilot contamination.
  Using~\eqref{eq:app_mr_term1},~\eqref{eq:app_mr_term2}, and~\eqref{eq:app_mr_term3} in~\eqref{eq:app_mr_start}, we conclude that
  \begin{IEEEeqnarray}{lCl}
    \!\!\!y_1^{\mathrm{ul}}[k] &\asymp &  \frac{1}{M}\tr({\bf{\Phi}}_1) x_1^{\mathrm{ul}}[k], \text{ if } \vecphi_1 \ne  \vecphi_2 \; \text{with}\;\vecphi_1^{\Htran} \vecphi_2 = 0, \! \!\nonumber\\ \\
      \!\!\! y_1^{\mathrm{ul}}[k] &\asymp & \frac{\tr({\bf{\Phi}}_1) x_1^{\mathrm{ul}}[k]+ \tr ({\bf \Upsilon}_{12}) x_2^{\mathrm{ul}}[k]}{M},  \text{ if } \vecphi_1 = \vecphi_2.\!\!\IEEEeqnarraynumspace
  \end{IEEEeqnarray}
  This implies that, in the absence of pilot contamination, the input-output relation becomes that of a deterministic noiseless channel as $M\to \infty$, while it converges to that of an \gls{awgn} channel  with transmit power $\lim_{M\to\infty}[ \frac{1}{M}\tr({\bf{\Phi}}_1) ]^2 $ and noise variance $\lim_{M\to\infty} [\frac{1}{M}\tr ({\bf \Upsilon}_{12})]^2$ when the two UEs use the same pilot sequence. 
  Note also that $g\asymp \widehat{g}\asymp \frac{1}{M}\tr({\bf{\Phi}}_1)$ where $g$ and $\widehat{g}$ are defined in~\eqref{eq:rcus_tail_simo_ul}.
  The desired result then follows from~\eqref{eq:rcus_noisefree}.

\section*{Appendix D - Proof of Theorem~\ref{thm:MMSE_asymptotic}}\label{app:M-MMSE_asymp}
We only consider the case in which the two \glspl{ue} use the same pilot sequence, i.e., $\vecphi_1 = \vecphi_2$.
By applying the matrix inversion lemma (see, e.g.,~\cite[Lem.~4]{Sanguinetti18}) we can rewrite \eqref{eq:MMSE_combiner} as
\begin{IEEEeqnarray}{lCl}
{\bf u}_1^{\rm{MMSE}} &=& \left(\sum\limits_{i=1}^2\widehat{\vech}_i\widehat{\vech}_i^{\Htran}  + {\bf Z}\right)^{-1}\widehat{\vech}_1\\ 
&=&  \frac{1}{1+ \gamma_1^{\text{ul}}} \left(\widehat{\vech}_2\widehat{\vech}_2^{\Htran}  + {\bf Z}\right)^{-1}\widehat{\vech}_1 \label{eq:MMSE_combiner_v1}
\end{IEEEeqnarray}
where we have set $\gamma_1^{\text{ul}} = \herm{\widehat{\vech}_1}\left(\widehat{\vech}_2\herm{\widehat{\vech}_2} + {\bf Z}\right)^{-1}\widehat{\vech}_1 $.
Substituting~\eqref{eq:MMSE_combiner_v1} into~\eqref{eq:simo_channel_ul} we obtain, after multiplying and dividing each term by $M$,
\begin{IEEEeqnarray}{lCl}
y_1^{\text{ul}}
  &=&  \frac{1}{\frac{1}{M} + \frac{\gamma_1^{\text{ul}}}{M}}\frac{1}{M}\herm{\widehat{\vech}_1}\left(\widehat{\vech}_2\herm{\widehat{\vech}_2} + {\bf Z}\right)^{-1} \!\!\!\vech_1 x_1^{\text{ul}} \nonumber\\
  &&{} + \frac{1}{\frac{1}{M} + \frac{\gamma_1^{\text{ul}}}{M}}\frac{1}{M}\herm{\widehat{\vech}_1}\left(\widehat{\vech}_2\herm{\widehat{\vech}_2} + {\bf Z}\right)^{-1} \!\!\!\vech_2  x_2^{\text{ul}}\nonumber\\
  &&{}+ \frac{1}{\frac{1}{M} + \frac{\gamma_1^{\text{ul}}}{M}}\frac{1}{M} \herm{\widehat{\vech}_1}\left(\widehat{\vech}_2\herm{\widehat{\vech}_2} + {\bf Z}\right)^{-1}\!\!\!\vecz^{\text{ul}}.\label{eq:MMSE_IO}
\end{IEEEeqnarray}
We begin by considering the first term. Under Assumption~\ref{assumption_1} and using~\cite[Lem.~3]{Sanguinetti18}, we obtain
\begin{align}
\frac{1}{M}\herm{\widehat{\vech}_1}\left(\widehat{\vech}_2\herm{\widehat{\vech}_2} + {\bf Z}\right)^{-1} \!\!\!\vech_1\asymp\frac{\gamma_1^{\text{ul}} }{M}
\end{align}
since $\vech_1 =\widehat{\vech}_1 + \widetilde{\vech}_1 $ with $\widetilde{\vech}_1$ being independent from $\widehat{\vech}_1$ and $\widehat{\vech}_2$.
 We note that, under Assumptions~\ref{assumption_1} and~\ref{assumption_2}, $\frac{\gamma_1^{\text{ul}}}{M}$ converges to a finite value as $M\to \infty$~\cite[App. B]{Sanguinetti18}. 
 This ensures that
\begin{align}
 \frac{\frac{\gamma_1^{\text{ul}}}{M}}{\frac{1}{M} + \frac{\gamma_1^{\text{ul}}}{M}} x_1^{\text{ul}} \asymp x_1^{\text{ul}}.
\end{align}
By applying~\cite[Lem.~5]{Sanguinetti18} to the second term in~\eqref{eq:MMSE_IO}, we obtain
\begin{IEEEeqnarray}{lCl}
  \IEEEeqnarraymulticol{3}{l}{
 \frac{1}{\frac{1}{M} + \frac{\gamma_1^{\text{ul}}}{M}}\frac{1}{M}\herm{\widehat{\vech}_1}\left(\widehat{\vech}_2\herm{\widehat{\vech}_2} + {\bf Z}\right)^{-1} {\vech}_2} \nonumber\\* \quad
 &=& \frac{1}{\frac{1}{M} + \frac{\gamma_1^{\text{ul}}}{M}}\Biggl(\frac{1}{M}\herm{\widehat{\vech}_1}{\bf Z}^{-1}{\vech}_2 \nonumber\\
 &&\qquad\qquad\qquad {}- \frac{\frac{1}{M}\herm{\widehat{\vech}_1}{\bf Z}^{-1}\widehat{\vech}_2\frac{1}{M}\herm{\widehat{\vech}_2}{\bf Z}^{-1}{\vech}_2}{\frac{1}{M} + \frac{1}{M} \herm{\widehat{\vech}_2}{\bf Z}^{-1}{\widehat{\vech}_2}} \Biggr).\label{eq: interference}\IEEEeqnarraynumspace
\end{IEEEeqnarray}
Under Assumption~\ref{assumption_1} and using~\cite[Lem.~3]{Sanguinetti18}, we have that\footnote{Under Assumption~\ref{assumption_1}, ${\bf Q}^{-1} {\bf R}_i{\bf Z}^{-1}{\bf R}_k$ has uniformly bounded spectral norm, which can be proved using in \cite[Lem.~4]{Sanguinetti18}.}, as ${M \to \infty}$, 
\begin{align}\label{eq:beta_11}
\frac{1}{M}\widehat{\bf h}_1^{\Htran}{\bf Z}^{-1}{\bf h}_2 &\asymp \frac{1}{M}\widehat{\bf h}_1^{\Htran}{\bf Z}^{-1}\widehat{\bf h}_2 \asymp \frac{1}{M}\tr ({\bf \Upsilon}_{12} {\bf Z}^{-1} ) \triangleq \beta_{12}\\
\label{eq:beta_22}
\frac{1}{M}\widehat{\bf h}_2^{\Htran}{\bf Z}^{-1}\widehat{\bf h}_2 &\asymp\frac{1}{M}\tr ( {\bf \Phi}_{2}{\bf Z}^{-1} ) \triangleq \beta_{22}.
\end{align}
Substituting~\eqref{eq:beta_11} and~\eqref{eq:beta_22} in~\eqref{eq: interference} and using Assumption~\ref{assumption_2}, we conclude that
\begin{IEEEeqnarray}{lCl}
 \frac{\frac{1}{M}}{\frac{1}{M} + \frac{\gamma_1^{\text{ul}}}{M}} \herm{\widehat{\vech}_1}\left(\widehat{\vech}_2\herm{\widehat{\vech}_2} + {\bf Z}\right)^{-1} {\vech}_2 &\asymp&  \frac{M}{\gamma_1^{\text{ul}}} \left(\beta_{12} - \frac{\beta_{12}\beta_{22}}{\beta_{22}}\right)\nonumber \\ 
 &=& 0
\end{IEEEeqnarray}
since $\frac{\gamma_1^{\text{ul}}}{M}$ converges to a finite value as $M\to \infty$ \cite[App. B]{Sanguinetti18}.
For the third term in \eqref{eq:MMSE_IO}, we have
\begin{IEEEeqnarray}{lCl}
\IEEEeqnarraymulticol{3}{l}{
 \frac{1}{\frac{1}{M} + \frac{\gamma_1^{\text{ul}}}{M}}\frac{1}{M}\herm{\widehat{\vech}_1}\left(\widehat{\vech}_2\herm{\widehat{\vech}_2} + {\bf Z}\right)^{-1} \!\!\!\vecz^{\text{ul}}}\nonumber\\*\quad 
 &=&   \frac{1}{\frac{1}{M} + \frac{\gamma_1^{\text{ul}}}{M}} \Biggl(\frac{1}{M}\herm{\widehat{\vech}_1}{\bf Z}^{-1}\vecz^{\text{ul}} \nonumber\\ 
 &&\qquad\qquad\qquad {} - \frac{\frac{1}{M}\herm{\widehat{\vech}_1}{\bf Z}^{-1}\herm{\widehat{\vech}_2}\frac{1}{M}\herm{\widehat{\vech}_2}{\bf Z}^{-1}\vecz^{\text{ul}}}{\frac{1}{M} + \frac{1}{M} \herm{\widehat{\vech}_2}{\bf Z}^{-1}{\widehat{\vech}_2}} \Biggr).\IEEEeqnarraynumspace
\end{IEEEeqnarray}
Under Assumption~\ref{assumption_1} and using~\cite[Lem.~3]{Sanguinetti18}, we have that, as ${M \to \infty}$,
\begin{align}
\frac{1}{M}\herm{\widehat{\vech}_1}{\bf Z}^{-1}\vecz^{\text{ul}}  \asymp0 \quad \text{and} \quad
\frac{1}{M}\herm{\widehat{\vech}_2}{\bf Z}^{-1}\vecz^{\text{ul}}  \asymp0
\end{align}
where we have used that $({\widehat{\vech}_1} , {\widehat{\vech}_2})$ and $\vecz^{\text{ul}}$ are independent. Therefore, we have that
\begin{align}
 \frac{1}{\frac{1}{M} + \frac{\gamma_1^{\text{ul}}}{M}}\frac{1}{M}\herm{\widehat{\vech}_1}\left(\widehat{\vech}_2\herm{\widehat{\vech}_2} + {\bf Z}\right)^{-1} \vecz^{\text{ul}}\asymp 0.
\end{align}
Combining all the above results, we conclude that $y_1^{\text{ul}}
  \asymp x_1^{\text{ul}}$. This implies that, as $M\to\infty$, the input-output relation~\eqref{eq:MMSE_IO} converges to that of a deterministic noiseless channel. 
The desired result then follows from~\eqref{eq:rcus_noisefree}. 

\bibliographystyle{IEEEtran}

\end{document}